\documentclass[11pt]{article}
\pdfoutput=1
\usepackage{jheppub}
\usepackage{tabu}
\usepackage[vcentermath]{youngtab}
\usepackage[usenames,dvipsnames,table]{xcolor}
\usepackage{graphicx,amsmath,amssymb,amsthm,multirow,array,bm,bbm,esint}
\usepackage[mathscr]{eucal}
\usepackage[bbgreekl]{mathbbol}  
\usepackage{epsf,amsfonts}
\usepackage{slashed}
\usepackage[numbers,sort&compress]{natbib}
\usepackage{minibox}
\usepackage{rotating}
\usepackage{pdflscape}
\usepackage{array,tikz-cd}
\usepackage{subfigure}

\usepackage{soul,xcolor}
\setstcolor{red}

\usepackage{slashed}

\definecolor{rust}{rgb}{0.8,0.2,0.2}

\newcommand{\prn}[1]{\left ( #1 \right )}
\newcommand{\brk}[1]{\left [ #1 \right ]}

\newcommand{\Tr}[1]{\hbox{Tr}\left(#1\right)}

\newcommand{\vev}[1]{\langle #1 \rangle}
\def\ket#1{\mid  \! #1  \rangle}
\def\bra#1{\langle  #1 \! \mid}

\def\kett#1{\mid\!\mid \! #1 \rangle\!\rangle}
\def\braa#1{\langle\!\langle #1 \! \mid\!\mid}

\def\weyl#1{: #1 :{}_{_{\!W}}}

\newcommand{\rhoi}{\hat{\rho}_{\text{initial}}}

\newcommand{\Op}[1]{\mathbb{#1}}
\newcommand{\OpH}[1]{\widehat{\mathbb{#1}}}

\newcommand{\SKR}[1]{\mathbb{#1}_{\skR}}
\newcommand{\SKL}[1]{\mathbb{#1}_{\skL}}

\newcommand{\SKAv}[1]{\mathbb{#1}_{{av}}}
\newcommand{\SKRel}[1]{\mathbb{#1}_{{dif}}}
\newcommand{\SKDif}[1]{\mathbb{#1}_{{dif}}}

\newcommand{\SKG}[1]{\mathbb{#1}_{_G}}
\newcommand{\SKGb}[1]{\mathbb{#1}_{_{\overline{G}}}}

\newcommand{\gh}[1]{\text{gh}(#1)}

\newcommand{\QSK}{\mathcal{Q}_{_{SK}}}
\newcommand{\QSKb}{\overline{\mathcal{Q}}_{_{SK}}}

\newcommand{\gradcomm}[2]{ \brk{ #1, #2 }_{\scriptscriptstyle \pm} }

\newcommand{\avB}{ {\sf a} }
\newcommand{\ghA}{ {\sf g}_\avB }
\newcommand{\ghbA}{ {\bar {\sf g}}_\avB }
\newcommand{\ggbA}{ {\sf a}_{\sf{g}\bar{\sf g}}}

\newcommand{\difB}{ {\sf f}_{\sf d} }
\newcommand{\difT}{ {\sf d} }
\newcommand{\ghD}{ {\sf g}_\difT }
\newcommand{\ghbD}{ {\bar {\sf g}}_\difT }
\newcommand{\ggbD}{ {\sf d}_{\sf{g}\bar{\sf g}}}

\newcommand{\thb}{{\bar{\theta}} }
\newcommand{\thetab}{\bar{\theta} }
\newcommand{\SF}[1]{\mathring{#1}}

\newcommand{\psib}{\overline{\psi}}
\newcommand{\Psib}{\overline{\Psi}}

\makeatletter
  \newcommand\Ttiny{\@setfontsize\Ttiny{1pt}{2}}
\makeatother

\newcommand{\Cref}{{\sf C}}

\newcommand{\skR}{\text{\tiny R}}
\newcommand{\skL}{\text{\tiny L}}

\title{Schwinger-Keldysh superspace in quantum mechanics}

\author[a]{Michael Geracie}
\author[b]{\!, Felix M. Haehl}
\author[c]{\!, R.\ Loganayagam}
\author[c]{\!, Prithvi Narayan}
\author[a]{\!,\\ David M.~Ramirez}
\author[a]{\!, Mukund Rangamani}

\affiliation[\,a]{
Center for Quantum Mathematics and Physics (QMAP),  \\
Department of Physics, University of California, Davis, CA 95616 USA.}
\affiliation[\,b]{Department of Physics and Astronomy, University of British Columbia,\\
6224 Agricultural Road, Vancouver, B.C.\ V6T 1Z1, Canada.}
\affiliation[\,c]{International Centre for Theoretical Sciences (ICTS-TIFR), \\
Shivakote, Hesaraghatta Hobli, Bengaluru 560089, India.}

\emailAdd{michael.geracie@gmail.com}
\emailAdd{f.m.haehl@gmail.com}
\emailAdd{nayagam@gmail.com}
\emailAdd{prithvi.narayan@gmail.com}
\emailAdd{dramir@ucdavis.edu }
\emailAdd{mukund@physics.ucdavis.edu}

\abstract{
We examine, in a quantum mechanical setting, the Hilbert space representation of the BRST symmetry associated with Schwinger-Keldysh path integrals. This  structure had been postulated to encode important constraints on influence functionals in coarse-grained systems with dissipation, or in open quantum systems. Operationally, this entails uplifting the standard Schwinger-Keldysh two-copy formalism into superspace by appending BRST ghost degrees of freedom. These statements were previously argued at the level of the correlation functions. We provide herein a complementary perspective by working out the Hilbert space structure explicitly.  Our analysis  clarifies two crucial issues not evident in earlier works:  firstly, certain background ghost insertions necessary to reproduce the correct Schwinger-Keldysh correlators arise naturally.  Secondly, the Schwinger-Keldysh difference operators are systematically dressed by the ghost bilinears, which turn out to be necessary to give rise to a consistent operator algebra. We also elaborate on the structure of the final state (which is BRST closed) and the future boundary condition of the ghost fields.
}

\begin{document}
\maketitle


\section{Introduction}
\label{sec:intro}
 
 The Schwinger-Keldysh formalism \cite{Schwinger:1960qe,Keldysh:1964ud,Feynman:1963fq} allows one to compute real-time observables in either closed or open quantum systems prepared initially in  a (w.l.o.g) mixed state. The basic idea behind the construction involves working with either a complex time contour that doubles back onto the starting configuration, or equivalently working with a double copy of the original system. The rationale for the doubling can be ascribed to the entanglement inherent in the initial state as is clear from the explicit path integral arguments of \cite{Feynman:1963fq}. The formalism is well developed and has been applied to many interesting physical systems over the years; see e.g., \cite{Chou:1984es} for a comprehensive review.

One central question that has remained unclear in the formalism is the nature of interactions between the two copies of the doubled system. These contributions, which were called influence functionals in \cite{Feynman:1963fq}, should obey some constraints reflecting the underlying quantum evolution. For closed quantum systems the constraints would encode microscopic unitarity, while for open quantum systems these would arise from evolution engendered by completely positive trace preserving quantum operations. In particular, such constraints on influence functionals are imperative if we are interested in integrating out a subset of degrees of freedom in the Schwinger-Keldysh functional integral, as we would for instance in the context of the renormalization group to extract the low energy effective dynamics.

Inspired by the structure of certain Ward identities that pertain for Schwinger-Keldysh observables \cite{Chou:1984es,Weldon:2005nr}, some of us argued in \cite{Haehl:2015foa} that the Schwinger-Keldysh construction should be interpreted not in terms of a two-copy system, but rather as a quadrupled system with a pair of topological BRST/anti-BRST symmetries $\{\QSK, \QSKb\}$ acting naturally. The basic idea was to append, to the doubled system, opposite Grassmann parity ghost systems. Should the original quantum system consist of only bosonic degrees of freedom, one would add a pair of Grassmann odd ghosts. 
Concurrently,  \cite{Crossley:2015evo} also argued for a BRST symmetry in the Schwinger-Keldysh construction. In both cases the idea of introducing the BRST symmetries was to constrain the low-energy dynamics and obtain an effective action for non-linear dissipative systems, specifically actions for relativistic hydrodynamics. Such actions were constructed independently in  \cite{Crossley:2015evo} and \cite{Haehl:2015uoc}, which explicitly exploit this Schwinger-Keldysh BRST (henceforth SK-BRST) symmetry (along with some additional structure arising from thermal density matrices and the KMS condition).

The construction of  \cite{Haehl:2015foa} has been further elaborated upon in \cite{Haehl:2016pec}, where formal arguments were given regarding the nature of the BRST symmetry and its action on the operator algebra of the quantum system. It was argued there that the natural way to view the Schwinger-Keldysh construction is in terms of a superspace with two Grassmann-odd directions (parameterized by say $\theta$ and $\thb$). The BRST symmetries act as super-derivations on the extended operator super-algebra, and it was also shown how to recover the Ward identities and fix the correlators of  ghost partners.

Likewise, \cite{Jensen:2017kzi} provide some additional discussion on the construction of \cite{Crossley:2015evo} (they also employ a superspace description similar to \cite{Haehl:2016pec}). 
It is worth noting that while \cite{Haehl:2015foa,Haehl:2016pec} demand a pair of BRST symmetries, the works of \cite{Crossley:2015evo,Jensen:2017kzi} argue for a single BRST supercharge (at least before introducing further constraints from thermality). We will here work with two supercharges which are naturally CPT conjugates of each other and refer the reader to \cite{Haehl:2017zac} for comments on the relative similarities/differences of the two approaches.\footnote{ Much of the focus of \cite{Crossley:2015evo,Jensen:2017kzi} lies only in the case of near-thermal density matrices, where one has to additionally account for the KMS condition. Our present discussion is general and not restricted to thermal states. For comments specific to thermal states please refer to \cite{Haehl:2015foa,Haehl:2016pec,Haehl:2016uah} for additional embellishments on the current discussion. It is also worth noting that the analysis of near-thermal systems in all these works requires an embedding into a superspace with two-Grassmann odd directions, though the origins for this structure are motivated very differently in \cite{Haehl:2015foa} and \cite{Crossley:2015evo}.}
We also note that  the constraints on influence functionals obtained by  explicit renormalization of an open $\phi^4$ theory \cite{Avinash:2017asn} are consistent with the Schwinger-Keldysh BRST charges posited in \cite{Haehl:2015foa,Haehl:2016pec}.\footnote{ To be clear, the discussion in \cite{Avinash:2017asn} strictly speaking only requires that the renormalized open $\phi^4$ theory admit an extension to include the aforementioned BRST structure. They show that the constraints they derive from a standard Schwinger-Keldysh doubled formalism can be derived much more simply by positing the action of BRST charges $\{\QSK, \QSKb\}$ that we espouse, together with a specific BRST allowed form of the  ghost action (and assuming further that the ghosts decouple in loops).  }

The formal discussions of these earlier works leave several questions unanswered. To enumerate a few salient ones:
\begin{itemize}
\item The SK-BRST symmetries  $\{\QSK, \QSKb\}$ were posited to act canonically on the extended operator super-algebra. In most quantum systems, we usually have a tendency to differentiate between simple/fundamental fields, and composite operators built from them, and it is unclear how the action of the SK-BRST charges on the former commutes with the OPE structure. This observation is independent of the number of BRST symmetries acting on the system and is equivalent to asking whether there is a Leibniz rule for  SK-BRST charges consistent with the OPE.
\item In checking the Schwinger-Keldysh Ward identities, and constructing the partner ghost correlators, \cite{Haehl:2016pec} had to argue for a background ghost insertion to soak up putative zero modes. Since the argument was at the operator level, a careful analysis of functional integral for zero modes was not made, and whilst the story was shown to be consistent, it was left unclear as to how these background ghosts arise.
\item The analysis of \cite{Haehl:2016pec} also presupposes the existence of a BRST closed final state; the details of its exact structure and the future boundary condition on the ghost modes were not fully explored.
\end{itemize}

The main aim of the current discussion is to try to clear up these loose ends and give a clean description of the Schwinger-Keldysh formalism including these SK-BRST symmetries. To illustrate various points without getting tangled up in details, we choose to work in the context of single-particle quantum mechanics, and moreover use the harmonic oscillator as our prime example to illustrate some important features of the construction. A clear advantage is that the operator algebra is now spanned by finitely many generators, which are the canonically conjugate variables of the system. Once we address the aforementioned questions in this primitive setting, we should then be able to make a general argument which would apply in other quantum systems (including QFTs). 

We find that there is ample freedom in how one can embed the Schwinger-Keldysh doubled formalism into an enlarged Hilbert space where our BRST symmetries act naturally. The initial state of our quantum system as well as the future boundary condition of the Schwinger-Keldysh construction are uplifted into this extended Hilbert space, albeit with some freedom. Along with this uplift, we also demonstrate how to uplift the operator algebra in a fashion consistent with the OPE structure. A novel feature of this discussion, which was not fully appreciated in \cite{Haehl:2016pec}, is the fact that the difference operators of the Schwinger-Keldysh formalism get dressed with BRST ghost bilinears to ensure that the OPE structure is sensible. We furthermore find that the quantum mechanical problem singles out Weyl ordering of operators; these are natural in the Schwinger-Keldysh construction owing to the fact that the temporal ordering is reversed between the forward and backward legs of the timefold contour.

The outline of the paper is as follows: In \S\ref{sec:skreview} we give a quick synopsis of background material relating to the  Schwinger-Keldysh formalism and the BRST symmetries we need for the discussion. In \S\ref{sec:SKHilbert} we then rephrase this discussion directly in terms of a Hilbert space picture and outline the necessary conditions we must satisfy when we extend the structure to include the BRST ghosts. In \S\ref{sec:QHO}
we demonstrate how these constraints can be satisfied in the simplest quantum mechanical setting: a quantum harmonic oscillator. Finally, we close in \S\ref{sec:discuss} with a discussion, indicating how the construction can be generalized to include interactions and go beyond single-particle quantum mechanics, and lay out some other interesting open questions.

\section{BRST symmetry in Schwinger-Keldysh: A review}
\label{sec:skreview}

The Schwinger-Keldysh generating functional which computes real time correlation functions in a specified (possibly mixed) initial state $\rhoi$, is
\begin{equation}
\mathscr{Z}_{SK}[\mathcal{J}_\skR, \mathcal{J}_\skL]  
	= \Tr{U[\mathcal{J}_\skR] \, \rhoi\,  (U[\mathcal{J}_\skL])^\dagger } .
\label{eq:skgen}
\end{equation}	
The basic idea behind this construction is that one wishes to be agnostic of the final state system when acted upon with sources. To ensure that one can probe the system, correlation functions are defined as matrix elements in the initial state, which requires that one evolves the system, inserts various operators, and then evolves back to the initial state (the formalism is hence sometimes referred to as the `in-in formalism').

One common way to interpret the Schwinger-Keldysh path integral is to view the integration contour as extending into complex time, where the forward and backward legs have infinitesimal separation in the imaginary direction. The forward evolving segment possesses background fields $\mathcal{J}_\skR$ while the backward evolving segment contains fields $\mathcal{J}_\skL$, corresponding to evolution according to $U[\mathcal{J}_\skR]$ and $U[\mathcal{J}_\skL]^\dagger$, respectively. Alternately, we can consider the forward and backward legs of the contour as independent evolutions and work with two copies of the original quantum system 
(indexed now by R and L) with matching boundary conditions at the turning point. From a calculational viewpoint the latter interpretation is often convenient and thus one naturally ends up working with two copies of the original system. In what follows, we will work with conventions of \cite{Haehl:2016pec}, where elements of the operator algebra of the quantum system of interest will be denoted with a hat $\OpH{O}$, while the Schwinger-Keldysh double-copy operators will be unhatted but subscripted, viz., $\SKR{O}$ and $\SKL{O}$, respectively.

Since in the Schwinger-Keldysh construction we can insert operators on either the L or R contours via functional differentiation, $\mathscr{Z}_{SK}[\mathcal{J}_\skR, \mathcal{J}_\skL] $ gives us access to a larger number of correlation functions than the standard single copy partition function \cite{Chou:1984es}, which generates time ordered expectation values.\footnote{ The number of contour $n$-point correlators is $2^n$, while there are only $2^{n-1}$ Schwinger-Keldysh ordered correlators. The latter count follows from the number of time-orderings of $n$ operators involving Heisenberg evolution with exactly one forward and one backward contour, a.k.a.\  1-timefold or  1-OTO. See  \cite{Haehl:2017qfl} for further details. } 
This implies that there should be various relations between the Schwinger-Keldysh contour correlators. These are captured by  simple rules (cf., \cite{Weldon:2005nr}):
\begin{itemize}
\item Any correlation function of an arbitrary number of difference operators, $\SKDif{O} \equiv \SKR{O} - \SKL{O}$, vanishes 
independent of the location of the insertions.
\item Correlators with a difference operator as the futuremost insertion vanish, i.e., the \emph{largest time equation} holds,
\begin{align}
	\left \langle \mathcal T_{_{SK}} \SKDif{O} (t) \prod_i \Op{O}_i (t_i)\right \rangle = 0 \,,
\end{align}
if $t>t_i$ for all $i$.
\end{itemize}
The first rule is of course a special case of the second.
These relations can be inferred directly from \eqref{eq:skgen} by noting that the Schwinger-Keldysh path integral involves the source deformed action
\begin{equation}
S_{SK} = S[\Phi_\skR] - S[\Phi_\skL] + \int d^dx\, \prn{\mathcal{J}_\skR \, \SKR{O} - \mathcal{J}_\skL \, \SKL{O}}\,.
\label{eq:Ssk}
\end{equation}	
By a basis rotation $\SKAv{O} = \frac{1}{2} (\SKR{O} + \SKL{O})$, the Lorentz signature source-operator coupling can be put in light-cone form: $\mathcal{J}_{dif} \, \SKAv{O} + \mathcal{J}_{av} \, \SKDif{O}$. Since the average sources couple to the difference operators, setting $\mathcal{J}_\skL = \mathcal{J}_\skR$ in \eqref{eq:skgen} suffices to generate difference operator insertions. At the same time, by unitarity of the evolution operator, the generating functional (\ref{eq:skgen}) collapses into the trace over the initial state
\begin{align}
	\mathscr{Z}_{SK}[\mathcal{J}, \mathcal{J}]  =1\,.
\end{align}

The vanishing of difference operator correlators, tantamount to a statement of unitarity, should be encapsulated as a general principle of the Schwinger-Keldysh construction.
It was therefore posisted  in \cite{Haehl:2015foa} and elaborated upon further in \cite{Haehl:2016pec} that  a useful way of viewing the Schwinger-Keldysh path integral is in terms of a quadrupled operator algebra with a topological BRST symmetry. As mentioned in \S\ref{sec:intro}, related  observations were also made in \cite{Crossley:2015evo} (see also \cite{Jensen:2017kzi}).

To wit, it was proposed that there are CPT-conjugate BRST charges $\QSK$ and $\QSKb$, satisfying a superalgebra
\begin{equation}
	\{ \QSK , \QSK \} =
	\{ \QSKb , \QSKb \} =
	\{ \QSK , \QSKb \} = 0,
\label{eq:SUSYalgebra}
\end{equation}	
which are engineered such that  the difference operators are BRST-descendants. That is,
\begin{equation}
\exists \; \SKG{O} \,, \SKGb{O} \, :  \quad  \SKDif{O} = - \gradcomm{\QSK}{\SKGb{O}} = \gradcomm{\QSKb}{\SKG{O}}
\label{eq:Exact} .
\end{equation}	
The nilpotency of $\QSK, \QSKb$ then implies that %
\begin{equation}
\gradcomm{\QSK}{\SKDif{O}}  = \gradcomm{\QSKb}{\SKDif{O}}   =0 \,.
\label{eq:Closed}
\end{equation}	
The operators $\SKG{O}$ and $\SKGb{O}$ carry opposite Grassmann statistics  relative to the original operator $\OpH{O}$ and have equal and opposite (conserved) ghost number. The BRST structure can be summarized by the (graded) commutation diagram
\begin{equation}
\begin{tikzcd}
&\SKAv{O} \arrow{ld}{\QSK} \arrow{rd}[below]{\QSKb}    &   \\
\SKG{O}\arrow{rd}{\QSKb} & & \SKGb{O} \arrow{ld}[above]{\!\!\!\!\!\!\!\!\!\!\!\!\!\!-\QSK}\\
&   \SKRel{O} &
\end{tikzcd}.
\label{eq:qskactionAvDif}
\end{equation}
In other words, the $\{\QSK, \QSKb\}$ action on the operator algebra is
\begin{equation}\label{eq:QSKdefKeld}
\begin{split}
\gradcomm{\QSK}{\SKAv{O}}&=  \SKG{O},\quad
\gradcomm{\QSK}{\SKG{O} } = 0 ,\quad
\gradcomm{\QSK}{\SKGb{O}} = - \SKRel{O} ,\quad
\gradcomm{\QSK}{\SKRel{O}} = 0\ , \\
\gradcomm{\QSKb}{\SKAv{O}}&=  \SKGb{O},\quad
\gradcomm{\QSKb}{\SKGb{O} } = 0 ,\quad
\gradcomm{\QSKb}{\SKG{O}} =  \SKRel{O} ,\quad
\gradcomm{\QSKb}{\SKRel{O}} = 0 \,.
\end{split}
\end{equation}

These transformation rules can be efficiently summarized by introducing superspace, in which operators are taken as functions of spacetime (just time in quantum mechanics) and two Grassmann odd coordinates, $\theta$ and $\thb(= \theta^\dagger)$, respectively. All operators in the theory get uplifted to super-operators, and the multiplet \eqref{eq:qskactionAvDif} can be  collected into a single superfield
\begin{align}\label{eq:superfield}
	\SF{\Op{O}} = \SKAv{O} + \thb \;\SKG{O} + \theta \; \SKGb{O}+ \thb \theta \;\SKRel{O} .
\end{align}
In superspace, the Schwinger-Keldysh supercharges then act as super-derivations
\begin{align}
	\QSK \sim \partial_\thb\,,
	&&\QSKb \sim \partial_\theta\,,
\end{align}
which satisfy the algebra \eqref{eq:SUSYalgebra}.  More generally, our notation follows conventions used in \cite{Haehl:2016pec}. In particular,   we use $\big|$ to denote the projection to the bottom component and abbreviate the average and difference operators as $\Op{O} \equiv \Op{O}_{av}, \tilde {\Op{O}} \equiv \Op{O}_{dif}$. That is,
\begin{equation}
\SF{\Op{O}} \big| \equiv \SF{\Op{O}} \big|_{\theta = \thb = 0} \,,\qquad \qquad \text{so that}
	\qquad \qquad \SF{\Op{O}} \big|  = \Op{O}_{av}= \Op{O} \,, 
	\qquad \partial_\theta \partial_\thb \SF{\Op{O}} \big| =\SKRel{O}  = \tilde{\Op{O}}\,.
\label{eq:notation}
\end{equation}	
%

\section{Schwinger-Keldysh in Hilbert space}
\label{sec:SKHilbert}

The preceding discussion introduced the Schwinger-Keldysh partition function in its familiar setting of the functional integral, which is usually the most practical for computational purposes. However, the entire construction reviewed in \S\ref{sec:skreview} can be formulated directly on Hilbert space. We will be working in the canonical formulation since it makes operator ordering issues more explicit and sheds light on the superspace structure. The remainder of the paper is focused on developing this approach. At the end of the day we will end up with an extended Hilbert space including ghosts, on which we have an explicit action of  $\QSK$ and $\QSKb$ as linear operators. 

We begin by rephrasing the two-copy interpretation of the Schwinger-Keldysh contour in a Hilbert space picture in \S\ref{sec:skdouble}. We will then describe how to extend this to a supersymmetric description in \S\ref{sec:require}. 

\subsection{States in the doubled Hilbert space}
\label{sec:skdouble}

We begin with the Schwinger-Keldysh partition function, written explicitly as a trace over a Hilbert space $\mathcal H$ with basis $\ket{i}$
\begin{equation}
\begin{split}
\mathscr{Z}_{SK}[\mathcal{J}_\skR, \mathcal{J}_\skL]  
	&= \Tr{U[\mathcal{J}_\skR] \, \rhoi\,  U[\mathcal{J}_\skL]^\dagger } \\ 
	& = \sum_{i,jk}  \rho_{jk} \bra{i} U_\skR \ket{j} \; \bra{k}  U^\dagger_\skL \ket{i} \,,
\end{split}
\label{eq:skgen1}
\end{equation}	
where we have abbreviated $U[\mathcal{J}_{\skR,\skL}] \equiv U_{\skR,\skL}$.

The system has been prepared in a possibly mixed initial state
\begin{align}
	\rhoi = \sum_{jk} \rho_{jk} \ket{j} \bra{k}\,,
\end{align}
 which is itself a state in the tensor product  $\mathcal{H} \otimes \mathcal{H}^*$. We will think of this as a pure state in the doubled Hilbert space using the Choi isomorphism.\footnote{ The Choi isomorphism, or what sometimes is referred to as the channel-state duality, or the Jamiolkowski-Choi isomorphism, is formally the statement that the any quantum channel can be equivalently represented as a state in a bipartite Hilbert space. While the idea is usually applied to quantum gates implementing operations, since the density matrix  is also an operator acting on the Hilbert space we find it natural to extend the terminology to apply to mixed states. We should also note that there are some operational distinctions, involving conjugations, etc., in the way various of these maps are defined on quantum operations; see \cite{Jiang:2013aa} for an overview of the literature.} Denote 
\begin{align}
 \ket{i} \bra{j} \  \;\; \longmapsto  \;\; \  \kett{ij}\,
\end{align}
These states form a basis for $\mathcal{H} \otimes \mathcal{H}^*$.\footnote{ For the sake of clarity, it is helpful to indicate  the Hilbert space index explicitly at the outset -- i.e., $ \ket{i_\skR} \bra{j_\skL}$ and $ \kett{i_\skR\, j_\skL}$, respectively. We refrain from using the labels in the text to avoid clutter; the $\mathcal{H}_\skR$ states precede those of  $\mathcal{H}_\skL$ states in both the bra and the ket of $\mathcal{H}_\skR \otimes \mathcal{H}_\skL^*$.}  The system thus begins in the state
\begin{equation}
\rhoi ~ \longmapsto ~ \kett{\rho_{_{SK}}} = \sum_{jk} \, \rho_{jk} \, \kett{j \, k}.
\label{eq:rsk}
\end{equation}	
Similarly, the trace is represented by the (un-normalized) maximally entangled state in $\mathcal{H} \otimes \mathcal{H}^*$
\begin{equation}
\kett{f_{_{SK}}} = \sum_i\, \kett{i\, i }\,.
\label{eq:fsk}
\end{equation}	
We can then re-write the Schwinger-Keldysh generating functional as a matrix element
\begin{align}
\sum_{i,jk}  \rho_{jk} \, \bra{i} U_\skR \ket{j} \; \bra{k}  U^\dagger_\skL \ket{i} 
  = \braa{ f_{_{SK}} } U_{_{SK}}  \kett{\rho_{_{SK}}} \,,
\end{align}
where we we have denoted the Schwinger-Keldysh propagator as
\begin{align}
U_{SK} = U[\mathcal{J}_\skR] U[\mathcal{J}_\skL]^\dagger=e^{ -i H_{SK} \,t }  ,\qquad 
H_{SK} \equiv H_\skR - H_\skL = \widehat{H} \otimes \Op{1} - \Op{1} \otimes \widehat{H} .
\end{align}
All told,
\begin{align}
\mathscr{Z}_{SK}[\mathcal{J}_\skR, \mathcal{J}_\skL]   = 
\sum \; \braa{f_{_{SK}}} U_{_{SK}}  \kett{\rho_{_{SK}}} \,.
\label{eq:skgen2}
\end{align}	
This is a trivial rewriting of \eqref{eq:skgen1}. Conceptually, however, one imagines starting from the initial state, $\rhoi$, evolving the kets forward, the bras backward, and subsequently evaluating the overlap with the (unnormalized) maximally entangled state.  	

Operators inserted on the right and left contours of the Schwinger-Keldysh path integral then enter the canonical formalism as acting on kets or bras respectively. Hence we denote
\begin{align}
	\SKR{A} = \OpH{A} \otimes \Op{1},
	&&\SKL{A} = \Op{1} \otimes \OpH{A} .
\end{align}
Here $\OpH{A}$ is a given operator on $\mathcal H$ and $\SKR{A}$, $\SKL{A}$ are operators on $\mathcal H \otimes \mathcal H^*$. We will always denote operators acting on $\mathcal{H}$ with a hat, and index operators with R, L subscripts to indicate whether they act from the right or the left on the density matrix in 
\eqref{eq:skgen1}.  

While seemingly trivial, there are some subtle issues with regards to operator ordering once we adopt this Choi map. Operators on the first and second Hilbert spaces in $\mathcal{H}\otimes \mathcal{H}^*$ act as
\begin{align}
 \SKR{A} \SKL{B} \ket{i} \bra{j}  =  \OpH{A} \ket{i} \bra{j} \OpH{B}  \;\; \longmapsto \;\; 
\SKR{A} \,  \SKL{B} \kett{ij}  = \OpH{A} \otimes \OpH{B} \kett{ij}  \,. 
\end{align}
Since L operators act on bras on the right, after the Choi  map, the algebra induced on operators on  $\mathcal{H} \otimes \mathcal{H}^*$ is  somewhat non-intuitive: products of operators on $\mathcal{H}^*$ are reversed. That is,
\begin{equation}
\begin{split}
\ket{i} \bra{j} \OpH{A}\OpH{B}
& \;\; \longmapsto \;\;  (\OpH{1}\otimes \OpH{B})( \OpH{1}\otimes \OpH{A}) \kett{ij}  = \SKL{B}\, \SKL{A} \,\kett{ij} .
\end{split}
\end{equation}
More succinctly,
\begin{align}
	\SKL{B} \SKL{A} = \SKL{(AB)}.
\end{align}
In general, given a string of R operators, the analogous string of L operators involves an order reversal.

\subsection{Superspace uplift and constraints}
\label{sec:require}

While this is the story for the standard Schwinger-Keldysh contour, we would like to ask whether we can extend this construction to an enlarged Hilbert space that will allow us to identify the ghost operators and topological symmetries. There are two independent sets of requirements for such an embedding. One involves constructing suitable states to represent the system in the extended Hilbert space. The other involves working out the correct multiplet structure of the super-operators (particularly for composite operators). It was assumed in \cite{Haehl:2016pec} that both of these could be done. This turns out to be true, but there are some important subtleties in the construction that become obvious in the Hilbert space picture and that we will try to flesh out here. In this section we will discuss the embedding in the extended Hilbert space. We turn to the more subtle issue of the supermultiplet structure in \S\ref{sec:QHOcomp}.

\paragraph{States in the extended Hilbert space:} 

In \cite{Haehl:2016pec}, the idea was to extend the operator algebra to a superalgebra by introducing opposite Grassmann parity ghost operators into the path integral.  In the Hilbert space picture, this amounts to extending the Hilbert space
\begin{align}
\mathcal{H} \otimes \mathcal{H}^* 
& \rightarrow
	 \mathcal{H} \otimes \mathcal{H}^* \otimes \mathcal{H}_{ghosts} \,, 
	 &&
\mathcal{H}_{ghosts} 
\equiv
	\mathcal{H}_{_G} \otimes \mathcal{H}_{_{\bar G}} \,.
\end{align}
The initial and final states need to be specified in the enlarged Hilbert space. We denote these as $\kett{\SF{f}}$ and $\kett{\SF{\rho}} $, which extend (\ref{eq:rsk}) and (\ref{eq:fsk}) into the quadrupled Hilbert space
\begin{equation}
\begin{split}
\kett{\SF{\rho}}  &= \rho_{ij\alpha \beta} \kett{i\, j \, \alpha \, \beta} ,\\
\kett{\SF{f}}  &= f_{ij\alpha \beta} \kett{i\, j \, \alpha \, \beta} .\\
\end{split}
\label{eq:fisuper}
\end{equation}

We demand that super-extension satisfies the following requirements:
\begin{enumerate}
\item $\kett{\SF{f}}$ is annihilated by $\QSK$ and $\QSKb$.
\item $\kett{\SF{f}}$ is a zero-energy eigenstate of the (extended) Hamiltonian $\mathscr{H}$.
\item $\kett{\SF{\rho}}$ and $\kett{\SF{f}}$ are selected so that correlation functions without ghost insertions reduce to those computed using the original generating functional \eqref{eq:skgen2}. This translates to the condition
\begin{equation}
\begin{split}
\braa{\SF{f}} \Op{O} \otimes \Op{1}_{ghosts}\kett{\SF{\rho}} &= \braa{f_{SK}}  \Op{O} \kett{\rho_{SK}}
\end{split}
\label{eq:ghostdecoupling}
\end{equation}
for any operator $\Op{O}$ on the doubled Hilbert space $\mathcal H \otimes \mathcal H^*$. For instance, $\langle\!\langle \SF{f}  \mid\!\mid \SF{\rho}\rangle\!\rangle =1$.
\end{enumerate}
Note that it is by no means obvious that these conditions can be satisfied. Demonstrating their consistency is the main goal of this paper. 

Of the above, condition 3 is most intuitive, since in the absence of ghost insertions the correlation functions should reduce to the ones computed in the doubled theory. 
Condition 1 is imposed so as to not break supersymmetry by our selection of the final state. In particular, we would like the largest time equation to arise from the $\QSK$ and $\QSKb$ exactness of difference operators.
Condition 2 constrains the final state so that the partition function on the now extended Hilbert space localizes after the future-most operator insertion. It is equivalent to requiring that $\braa{ \SF{f}} \mathcal U(t) ( \cdots ) \kett {\SF{\rho}}$ is bereft of any phase factors coming from unitary evolution.

The dynamics in the extended Hilbert space is dictated by  the Hamiltonian  
\begin{align}
\mathscr{H} = H_{SK} \otimes \Op{1}_{ghosts} + H_{ghosts}, \qquad \quad
 \mathscr{U} (t) = e^{- i \,\mathscr{H}\, t} ,
\end{align}
acting on the enlarged Hilbert space. Our aim is to keep this extension compatible with the supersymmetry so that
\begin{align}
	 [ \QSK, \mathscr{H}  ]= [  \QSKb ,\mathscr{H} ] =  0
\end{align}
when the sources are aligned $\mathcal J_\skR = \mathcal J_\skL$ (which we are always implicitly assuming). As we shall see $H_{ghosts}$ only has nontrivial action on the ghost Hilbert space for free theories, but will contain a nontrivial bosonic part in interacting theories.

Should such a super-extension exist, one can directly calculate all correlation functions, including those of the ghost partners $\SKG{O}$ and $\SKGb{O}$. In \cite{Haehl:2016pec} it was argued that one needs to admit a background ghost dressing into the correlation functions for these to be consistent. This was necessary to ensure no violations of the largest time equation. Clearly, with an explicit Hilbert space realization of the ghosts and supercharges at hand, we should be able to clarify the origin of the background ghosts. In \S\ref{sec:sliding} we find that they are simply the images of the initial state under $\QSK$ and $\QSKb$. That is, they are ghost partners of the density matrix itself 
\begin{align}
	\QSK \kett{\SF{\rho}},
	&&\QSKb \kett{\SF{\rho}}.
\end{align}

Finally, now that we have discussed how to incorporate ghosts in the canonical formulation of quantum mechanics, let's take a moment to pause and discuss our motivations. 
One might ask why we are adding ghosts, since the original theory does not possess them. In \cite{Haehl:2016pec} it was argued that ghosts arise by gauge fixing a field reparameterization symmetry. Here we remain agnostic on how ghosts arise in practice, but merely assume they arise in some formulation, in the hope that this will make the symmetries of the Schwinger-Keldysh partition function manifest. However this happens, the theory will then contain the original physical degrees of freedom plus ghosts, and the reader may take that as our starting point. For a system as simple as the quantum harmonic oscillator, we don't anticipate that this will teach us anything new about the system itself; our goal in this paper is rather to examine the consequences of these ghosts in detail in a setting where the proposal can be made precise. It should be clear from the standard mode decomposition of operators in quantum field theory that a consistent formalism for the quantum harmonic oscillator goes a long way towards applying the same techniques to interacting theories (see \S\ref{sec:interact}).

\paragraph{Super-operators and the OPE:} Implicit in the above is the idea that we take every element of the operator algebra and convert it into an element of the operator super-algebra. While this appears to be reasonable, the operator algebra is required to be associative under the OPE, and one should ask how the $\QSK$, $\QSKb$ action distributes across the OPE. This becomes an issue, as we shall later see, in quantum systems where the algebra is built from a fundamental set of operators (which could for example be simply the creation/annihilation operators for a fundamental field), and for composite operators more generally.

Let us first give an abstract description of the task at hand. Given two operators in the single-copy theory  $\OpH{A}$, $\OpH{B}$, we would construct composite operators  $:\OpH{AB}:_k$ by using the OPE to normal order terms. In the doubled Schwinger-Keldysh formalism, this picture continues to hold in the L and R segments independently. As such, we would then naively want to associate new super-partners to these composite operators by invoking the action of the SK-BRST charges. To wit, if we assume the super-structure for $\Op{A}, \Op{B}$ is given by the action in \eqref{eq:qskactionAvDif}, we hope that we can then compute the $\QSK, \QSKb$ action on composites. Naively we would be tempted to write:
\begin{equation}
\begin{split}
\gradcomm{\QSK}{\SKR{:AB:}} & 
	\stackrel{?}{=} \gradcomm{\QSK}{\SKR{A}} \, \SKR{B}  + \SKR{A}  \, \gradcomm{\QSK}{\SKR{B}}  \\
\SKG{:AB:}\ &
	\stackrel{?}{=}\  :\SKG{A} \SKR{B} + \SKR{A} \SKG{B}:
\end{split}
\label{eq:compoSK}
\end{equation}	
and equivalently for the L-operators. We have made implicit some difficulties in implementing this in explicit examples with the question mark. This is the second issue we have to address to give a prescription for the super-extension of the Schwinger-Keldysh formalism.

Without further ado, we now turn to tackle these questions in the setting of single-particle quantum mechanics.

\section{Quantum harmonic oscillator}
\label{sec:QHO}
We first explore the questions raised in \S\ref{sec:require} by considering a free quantum theory, and add interactions. Since  nothing comes simpler than a harmonic oscillator we begin our discussion in this context. This already turns out to involve all the complications that need to be overcome, so there is no reason to consider a more complicated theory at this point.

The standard harmonic oscillator  action can be easily adapted to the Schwinger-Keldysh functional integral, by considering the action (with sources switched off):
\begin{align}
S_{SK} = \frac{1}{2} \int dt \left( m\,  \dot x_\skR^2 -   m\,\omega^2 x_\skR^2 
	- m\,  \dot x_\skL^2+  m\,\omega^2 x_\skL^2 \right) .
\end{align}
We work in units where $m=\omega = \hbar = 1$ for simplicity.

Since the theory is given in terms of a fundamental field $x$, it is easy to check that the natural action of the SK-BRST charges takes the form
\begin{align}
&\gradcomm{\QSK}{ x_{\skR,\skL}} =\psi \,, 
&&  
 \gradcomm{\QSKb}{x_{\skR,\skL}} = \psib,  \nonumber \\
&\gradcomm{\QSKb}{\psi}  = x_\skR - x_\skL\,,  
&& 
\gradcomm{\QSK}{\psib } = -( x_\skR - x_\skL) \,.
\label{eq:QHOsk1}
\end{align}
Consequently, including the ghost sector, we have the explicit BRST invariant action
\begin{align}
S = \frac{1}{2}\,  \int dt \left( \dot x^2_\skR -  x^2_\skR - \dot x^2_\skL +  x^2_\skL 
	+ 2\, \dot {\psib} \dot \psi - 2\, \psib \psi \right) .
\end{align}

Passing to quantum mechanics, $x \mapsto \OpH{X}$, $\dot{x} \mapsto \OpH{P}$, $\psi \mapsto \Psi$, $\dot{\psi} \mapsto P_{\Psib}$ the Hamiltonian is
\begin{align}
H &= 
	\frac{1}{2}\left(P_\skR^2 + X_\skR^2 - P_\skL^2 - X_\skL^2 \right) + 
	  P_{\Psi}\, P_{\Psib} + \Psib\,\Psi,
\end{align}
with canonical commutation relations 
\begin{align}
 [ X_\skR , P_\skR ]= i \,,
	&&  [X_\skL , P_\skL ] = -i \,,
	&& \{\Psi , P_\Psi \} = i \,,
	&&\{ \Psib , P_{\Psib} \} =  - i \,.
\label{eq:xpcomm}	
\end{align}
Recall that $X_\skR = \OpH{X} \otimes \Op{1}, \; P_\skR = \OpH{P} \otimes \Op{1}$, and $X_\skL = \Op{1} \otimes \OpH{X},\; P_\skL =\Op{1} \otimes \OpH{P}$. The above is then consistent with $[\OpH{X} , \OpH{P} ] = i$.

The action of the supercharges $\{\QSK, \QSKb\}$  given in  \eqref{eq:QHOsk1}, suitably uplifted to the operator algebra, can then be implemented as follows

\begin{equation}
	\begin{tikzcd}
		&X_{\skR, \skL} \arrow{ld}{\QSK} \arrow{rd}[below]{\QSKb}    &   \\
		\Psi \arrow{rd}{\QSKb} & & \Psib \arrow{ld}[above]{\!\!\!\!\!\!\!\!\!\!\!\!\!\! -\QSK}\\
		&   X_\skR - X_\skL &
	\end{tikzcd}
\hspace{2cm}
	\begin{tikzcd}
		&P_{\skR,\skL} \arrow{ld}{\QSK} \arrow{rd}[below]{\QSKb}    &   \\
		P_{\Psib} \arrow{rd}{\QSKb} & & P_{\Psi} \arrow{ld}[above]{\!\!\!\!\!\!\!\!\!\!\!\!\!\! -\QSK}\\
		&   P_\skR - P_\skL &
	\end{tikzcd}	.
\end{equation}

This can be achieved by writing down an explicit operator representation for the supercharges
\begin{equation}
\begin{split}
\QSK &=
	  i \left( P_\skR - P_\skL \right) \Psi  - i  \left(X_\skR - X_\skL \right) P_{\Psib} ,  \\
\QSKb & =
	 i \left( P_\skR - P_\skL \right) {\Psib} - i \left( X_\skR - X_\skL \right)  P_\Psi \,,
\end{split}
\label{eq:QactXP}	 
\end{equation}
which act on the operator algebra by graded commutators.
These charges satisfy a Hermiticity condition: $\QSK^\dagger = \QSKb$, where we also take 
$\Psib = - \Psi^\dagger$.

It is helpful at this stage to introduce superspace, in which the supercharges $\{\QSK, \QSKb\}$ act as super-derivations $\{\QSK \sim \partial_{\thb}, \ \QSKb \sim \partial_\theta\}$. Following \cite{Haehl:2016pec} we pick Grassmann odd coordinates $\theta$ and $\thb$ with non-zero ghost number (normalized such that $\gh{\theta} = +1$ and $\gh{\thb} =-1$). The charge assignment is consistent with the Hermiticity condition: 
$\theta^\dagger = \thb$. 
One can then upgrade the operators to super-operators $\Op{O} \mapsto \SF{\Op{O}}$, e.g., 
\begin{equation}
\begin{split}
\SF{X} &= 
	\frac{1}{2} \left(X_\skR + X_\skL\right) +  \theta\, \Psib +  \thb \, \Psi  
	+ \thb \theta\, \left(X_\skR - X_\skL\right) ,
 \\
\SF{P} &= 
	\frac{1}{2} \left(P_\skR + P_\skL\right) +   \theta\, P_\Psi +  \thb \, P_{\Psib}  
	+ \thb \theta\, \left(P_\skR - P_\skL\right) .
\end{split}
\label{eq:XPsuper}
\end{equation}
We will often find it convenient to abbreviate the average and difference operators above as
\begin{equation}
X \equiv X_{av} = \frac{1}{2} \left(X_\skR + X_\skL\right)  \,, \qquad \tilde X \equiv X_{dif} = X_\skR - X_\skL \,.
\label{eq:avdif}
\end{equation}	
In particular, note that the $X$ without a hat refers to the average Schwinger-Keldysh operator in the doubled theory.

\subsection{Ladder super-operators and Hilbert space}
\label{sec:aadsuper}

The Hilbert space is constructed from the application of creation/annihilation operators. These naturally reside in super-operators
\begin{equation}
\begin{split}
\SF{a} &= a  +  \thb \, c + \theta  \, b + \thb  \theta \, d \, ,\\
\SF{a}^\dagger &= a^\dagger   -\thb\,  b^\dagger - \theta \, c^\dagger + \thb  \theta \, d^\dagger \,,
\end{split}
\label{eq:aadSF}
\end{equation}
where $a,d$ are complex Grassmann-even fields and $b,c$ are Grassmann-odd fields with ghost charge 
$\gh{b} = -1$ and $\gh{c} = 1$, respectively.\footnote{ Note that the  Grassmann statistics and Hermiticity conditions imply signs under conjugation, viz.,  $(\theta \, b)^\dagger = b^\dagger \, \theta^\dagger = b^\dagger \, \thb = -\thb\, b^\dagger$. This is responsible for our anti-Hermiticity condition $\Psib = -\Psi^\dagger$} Here we have passed to the average/difference basis, and the relation with the R/L operators is
\begin{equation}
a \equiv \frac{1}{2} \left(a_\skR + a_\skL\right)  \,, \qquad  d \equiv a_\skR - a_\skL \,.
\end{equation}
The position and momentum super-operators are related to these as usual
\begin{equation}
\SF{X} = \frac{1}{\sqrt{2}}\left(\SF{a} + \SF{a}^\dagger\right) \,, \qquad 
\SF{P} = \frac{i}{\sqrt{2}} \left(\SF{a}^\dagger - \SF{a} \right) .
\label{eq:}
\end{equation}	
For reference, this is equivalent to
\begin{align}
	&  X_R = \frac{1}{\sqrt{2}} \left( a^\dagger_R + a_R \right),
	&&  P_R = \frac{i}{\sqrt 2} \left( a^\dagger_R - a_R \right), \nonumber \\
	&  X_R = \frac{1}{\sqrt{2}} \left( a^\dagger_L + a_L \right),
	&&  P_R = \frac{i}{\sqrt 2} \left( a^\dagger_L - a_L \right), \nonumber \\
	&  \Psi = \frac{1}{\sqrt 2} \left( c - b^\dagger\right),
	&&  P_\Psi = - \frac{i}{\sqrt 2} \left( c^\dagger + b \right) , \nonumber \\
	&  { \bar \Psi} = - \frac{1}{\sqrt 2} \left( c^\dagger - b \right) ,
	&&  P_{\bar \Psi} = - \frac{i}{\sqrt 2} \left( c + b^\dagger \right) .
	\label{eq:raisingLoweringDef}
\end{align}
The commutation relations follow from \eqref{eq:xpcomm}: 
\begin{equation}
\begin{split}
	[a_\skR , a_\skR^\dagger ] &= 1 
\,, \qquad  
	[a_\skL , a_\skL^\dagger ] = -1 
\,, \qquad  
	\{ b ,b^\dagger \} = 1 
\,, \qquad  
	 \{ c ,c^\dagger \} = -1
\end{split}
\label{eq:crelaad}
\end{equation} 

As indicated earlier, the  supercharges act as derivations, leading to the following action on the creation/annihilation operators:
\begin{align}
& 	[ \QSK ,a ] =  c  \,, 
&& 	\{\QSK ,   b \} = -d \,,  
&& 	[ \QSK ,a^\dagger ] = - b^\dagger\,,
&&	\{ \QSK ,  c^\dagger \} =  d^\dagger  \,, \nonumber \\
& 	[ \QSKb ,a ] = b  \
&&	\{ \QSKb,  c \} = d  \,, 
&&	[\QSKb ,a^\dagger ]= - c^\dagger \,,  
&&	 \{\QSKb , b^\dagger \} = -d^\dagger \,.
\label{eq:Qad}
\end{align}
One can also check that the Hamiltonian for the system (after applying the Choi map) is given in terms of these creation/annihilation operators as 

\begin{equation}
\mathscr{H}=   a_\skR^\dagger \, a_\skR - a_\skL^\dagger \, a_\skL + b^\dagger \, b + c\, c^\dagger\,,
\label{eq:Hamaad}
\end{equation}
while the supercharges themselves can be expressed as
\begin{equation}
\begin{split}
\QSK  &= 
	- \left(a_\skR - a_\skL \right)  b^\dagger - \left(a_\skR^\dagger - a_\skL^\dagger \right)  c \,, \\
\QSKb &=
	 - \left(a_\skR - a_\skL\right)   c^\dagger  -  \left(a_\skR^\dagger - a_\skL^\dagger\right)  b    \,.
\end{split}
\label{eq:Qaad}
\end{equation}

The system has two $U(1)$ charges. Firstly there is a  global symmetry under which $a_\skR ,a_\skL, b,c$ are charge $+1$, while their conjugates   $a_\skR^\dagger,a_\skL^\dagger, b^\dagger, c^\dagger$ carry charge $-1$. There is also an $U(1)_R$ symmetry under which $a_\skR ,a_\skL$ and their conjugates are neutral while 
$\gh{c,b^\dagger} = 1, ~\gh{b,c^\dagger} =-1$.

The total Hilbert space of the system is the tensor product of the two oscillators (R and L) and a two-state system for the SK-BRST ghosts. Note that the L oscillators are inverted, so $a_\skL $ should be treated as a creation operator, while $a^\dagger_\skL$ is the annihilation operator, which is clear from \eqref{eq:crelaad}.  We find it convenient to pick the following basis: for  ${\cal H}_\skR, {\cal H}_\skL$ choose the usual number operator basis  
\begin{equation}
\begin{split}
	a_\skR \ket{m_\skR} &= \sqrt{m_\skR} \ket{ m_\skR-1} 
\,, \qquad 
	a_\skR^\dagger \ket{m_\skR} = \sqrt{m_\skR +1 } \ket{ m_\skR+1} \,,  \\
 	a_\skL^\dagger \ket{n_\skL} &= \sqrt{n_\skL} \ket{ n_\skL-1} 
\,,\qquad\quad
 	 a_\skL  \ket{n_\skL} = \sqrt{n_\skL+1} \ket{ n_\skL+1} \,.
\end{split}
\end{equation} 
This can also be derived from $a_\skR = a \otimes \Op{1}, a_\skL = \Op{1}\otimes a$ and the reversal of operator orderings that occurs in the second factor of the tensor product.
For the $b$ and $c$ oscillators, select a ground state $\ket{00}$ annihilated by both, $b \ket{00} = c \ket{00} = 0$. The ghost Hilbert space is then spanned by $\ket{00}$ and
\begin{equation}
\begin{split}
| 10 \rangle \equiv & \  b^\dagger  \ket{00}\,, \\
| 01 \rangle \equiv & \    c^\dagger \ket{00} \,,\\
| 11 \rangle \equiv &\  b^\dagger c^\dagger \ket{00}\,.
\end{split}
\end{equation}
One can furthermore check that for $\alpha = 0,1$
\begin{align}
& 
	b \,| 0 \alpha \rangle = b^\dagger \,| 1 \alpha \rangle = 0 \,,
	&&c \,| \alpha0 \rangle = c^\dagger \,| \alpha 1 \rangle = 0 \,,\\
&
	b^\dagger\,| 0  \alpha \rangle = \,| 1\alpha \rangle \,, 
	&&b \,| 1\alpha \rangle = \,| 0\alpha \rangle \,, \\
&
	c^\dagger \,| \alpha 0 \rangle = (-1)^\alpha \,| \alpha1 \rangle \,, 
	&&c \,| \alpha 1 \rangle = (-1)^{\alpha+1}\,| \alpha 0 \rangle  \,.
\end{align}
The inner product on the ghost Hilbert space may be found in Appendix \ref{app:InnerProduct}.
Therefore the total Hilbert space after applying the Choi map can be decomposed  as
\begin{equation}
\begin{split}
\mathcal{H} &= 
	\mathcal{H}_\skR \otimes \mathcal{H}_\skL^* \otimes \mathcal{H}_b \otimes  \mathcal{H}_c \\
&=  	
	\text{span}\bigg\{
		\kett{ij \alpha \beta} \equiv 
		|i \rangle_\skR \otimes | j \rangle_\skL \otimes | \alpha \rangle_b \otimes | \beta \rangle_c \quad \bigg|
	\quad i,j \in {\mathbb Z}_+ \,,\  \alpha,\beta \in \{0,1\}\bigg\}\,.
\end{split}
\end{equation}
Dynamics on this Hilbert space is dictated by the Hamiltonian \eqref{eq:Hamaad} which acts as
\begin{equation}
\mathscr{H} \kett{ij\alpha \beta}  = \left(i-j + \alpha + \beta - 1 \right) \kett{ij\alpha \beta}\,.
\label{eq:}
\end{equation}	
%

\subsection{Composite operators}
\label{sec:QHOcomp}

Now that we have explicit expressions for the supercharges, we can ascertain how they act on composite operators. One might a-priori be tempted to posit that an operator of the form 
$X_{\skR}^n - X_{\skL}^n$ should be the top component of a superfield since it is a difference operator.  It is however easy to check that these are not $\QSK, \QSKb$ closed:
\begin{equation}
\begin{split}
[\QSK, X_\skR^n - X_\skL^n ] &= n \, \left(X_\skR^{n-1} - X_\skL^{n-1}\right) \, \Psi 
\,, \\
[\QSKb, X_\skR^n - X_\skL^n ] &= n \, \left(X_\skR^{n-1} - X_\skL^{n-1}\right) \, \Psib \,.\end{split}
\label{eq:Qcomp1}
\end{equation}	
Our plan is to remedy this by adding a suitable ghost-dressing, that is, we need keep track of ghost contributions to difference operators not considered in \cite{Haehl:2016pec}. However, this dressing should be incorporated without affecting physical  correlators.

The simplest way to proceed is to pass to superspace and construct super-operators satisfying certain requirements.  It will be useful to  introduce for any super-operator $\SF{\mathcal{O}}$ a deformation parameterized by a number $\zeta \in [-1,1]$, which shifts the bottom component by an amount proportional to the top component. Let
\begin{equation}
\SF{\mathcal{O}}_\zeta = \SF{\mathcal{O}} + \frac{\zeta}{2} \; \partial_\theta \partial_{\thb} \SF{\mathcal{O}} = 
	\SF{\mathcal{O}} + \frac{\zeta}{2}\, \tilde{\mathcal{O}}
\label{eq:Ozeta}
\end{equation}	
so that $\SF{\mathcal{O}}_{\zeta =1}\big| = \mathcal{O}_\skR $ and $\SF{\mathcal{O}}_{\zeta = -1}\big| = \mathcal{O}_\skL$.

Given the many composite operators in the theory, one can in principle construct many different super-multiplets. Let us enumerate a few salient multiplets in the extended operator super-algebra that will turn out to be sufficient for a comprehensive understanding of the structures of correlators. 

\paragraph{1. The product multiplet:} This multiplet is simplest to construct. Simply take suitable products of fundamental superfields. Starting with the position and momentum super-operators we could for example write:
\begin{equation}
\begin{split}
\SF{\Pi}_{m,n} \equiv \SF{X}^m  \SF{P}^n 
\equiv	 \left(X + \theta\, \Psib + \thb \, \Psi  + \thb \theta\, \tilde{X}\right)^m\, \,  
		\left(P + \theta\, P_\Psi + \thb \, P_{\Psib}  + \thb \theta\, \tilde{P}\right)^n  \,.
\end{split}	
\label{eq:prodmul}
\end{equation}	
 While simple, product multiplets will not play much of a role in our discussion.

\paragraph{2. The difference multiplet:} This is the multiplet we seek. Our plan is to engineer the top component to be a difference operator modulo additional ghost terms. Before doing so, we have to face up to the operator ordering issue seriously, since the sequence of operators on the left have to be reversed. 

Given a composite operator $\OpH{X}^m \OpH{P}^n$ in the single-copy description, the corresponding difference operator is actually $X_\skR^m \, P_\skR^n - P_\skL^n \, X_\skL^m$. We can choose to normal order any (post-Choi) operator as  $X_\skR^p\, P_\skR^q \, P_\skL^r\, X_\skL^s$ thus keeping track of the reversal explicitly. In principle, there is a straightforward algorithmic way of constructing the requisite ghost corrections to these difference operators; however, we found it cumbersome to implement generally.\footnote{ For instance, to construct the difference operator with Grassmann even part being $X_\skR^m \, P_\skR^n - P_\skL^n \, X_\skL^m$, we write down operators $S_{m,n} = \sum_{p,q,r,s} \, c_{pqrs}^{m,n} \,X_\skR^p\, P_\skR^q \, P_\skL^r\, X_\skL^s $ and determine the coefficients  $c_{pqrs}^{m,n}$ such that $\{\QSKb,[\QSK, S_{m,n}]\} = X_\skR^m \, P_\skR^n - P_\skL^n \, X_\skR^m +\text{ghosts}$. While we were able is write recursion relations that generate the required solution, the general structure was not immediately transparent. We therefore focus primarily on Weyl ordered operators which we think are more natural both from the perspective of the Choi isomorphism and the Schwinger-Keldysh operator algebra. If desired, once the Weyl ordered composite operators are determined, the canonical commutation relations may be employed to infer the super-operator structure for the normal ordered ones. We have checked that this agrees with the aforementioned analysis for small values of $m,n$.  }

One way to circumvent the issue  is  by switching to a Weyl ordered basis. 
We will denote the Weyl ordering of an operator $\OpH{A}$ built from $\OpH{X}$'s and $\OpH{P}$'s as $\weyl{\OpH{A}}$. For instance,
\begin{align}
	\weyl{\OpH{X}\, \OpH{P}^2} = \frac 1 3 \left( \OpH{X} \,\OpH{P}^2 + \OpH{P} \,\OpH{X} \,\OpH{P} + \OpH{P}^2 \,\OpH{X} \right) \,.
\end{align}
Similar statements hold for the double copy operators. Since Weyl ordered operators are palindromic in the basic operator alphabet (e.g., $\OpH{X}$ and $\OpH{P}$ for the harmonic oscillator), we have complete symmetry between the L and R. This choice then renders operator ordering concerns moot.  Given its simplicity we adapt it in what follows. Note that in the functional integral the Weyl ordering of operators is achieved by evaluating position dependent terms at the mid-point in the usual discretization procedure (see eg., the discussion in \cite{Srednicki:2007qs}).

We construct the general Weyl ordered difference operator as the top component of the super-operator built from an integral of powers of $\SF{X}_\zeta + \alpha \, \SF{P}_\zeta$. The parameter $\alpha $ is a book-keeping device used to pick out terms with a fixed number of momentum operators (i.e., it allows us to discuss a one-parameter family of multiplets simultaneously). The expression $(\SF{X}_\zeta + \alpha \, \SF{P}_\zeta)^k$ is a sum of Weyl ordered products $\weyl{\SF X^k_\zeta \SF P^{k-i}_\zeta}$ with coefficient $\alpha^i$. Define the difference multiplets as 
\begin{equation}
\begin{split}
\SF{\mathcal{D}}_{k} 
&=
	 \int_{-1}^1\, \frac{d\zeta}{2}\ \left(\SF{X}_\zeta + \alpha \,  \SF{P}_\zeta \right)^k  \\
&=
	\mathcal{D}_{k} 
	+ \thb \; k\, \mathcal{D}_{k-1}\left(\Psi  +  \alpha \, P_{\Psib}\right) 
	+\theta  \; k\, \mathcal{D}_{k-1}\left(\Psib  +\alpha\, P_{\Psi}\right)
	+ \thb \theta \,\widetilde{\mathcal{D}}_k \\
\mathcal{D}_k 
& \equiv \int_{-1}^1\,  \frac{d\zeta}{2} \,\left( X + \alpha\, P  + \frac \zeta 2 \left(\tilde{X} + \alpha\, \tilde{P} \right)\right)^k
\end{split}
\label{eq:difmul}
\end{equation}	
It is easy to see that the top component of this multiplet contains the difference operator we seek
\begin{equation}
\begin{split}
\widetilde{\mathcal{D}}_k
&= 
	\int d\theta \, d\thb \, \SF{\mathcal{D}}_k   \nonumber \\
& =
 	\int_{-1}^1 \frac{d \zeta}{2} \, \sum_{m=0}^{k-1}\,\left(X_\zeta + \alpha\, P_\zeta\right)^m\, (\tilde{X} + \alpha \tilde{P}) \, 
 	\left(X_\zeta + \alpha\, P_\zeta\right)^{k-1-m}  \\
&  	
	\hspace{3cm}
	 + \,k(k-1) \, \mathcal{D}_{k-2} \left(\Psib \,\Psi + 
 	\alpha\,\left(P_{\Psi} \, \Psi + \Psib\, P_{\Psib} \right) + \alpha^2\, 	P_{\Psi}\, P_{\Psib} \right)  \\ 
 & = \left(X_\skR + \alpha \, P_\skR\right)^k - \left(X_\skL+ \alpha \, P_\skL \right)^k \\
 &  	\qquad +\, k(k-1) \, \mathcal{D}_{k-2} \left(\Psib \,\Psi + 
 	\alpha\,\left(P_{\Psi} \, \Psi + \Psib\, P_{\Psib} \right) + \alpha^2\, 	P_{\Psi}\, P_{\Psib} \right)  \,. 
\end{split}
\label{eq:Dtop}
\end{equation}	
We have used the canonical commutation relations to bring the ghost contributions to a canonical form although we could have left them in Weyl ordered form too (this is more convenient for later computations). 

The above formulas are convenient since they efficiently collect the components of a difference multiplet into a single superfield. However, they can be rather opaque. Equivalently, one can show that the bottom component of the difference multiplet with $m$ factors of $X$ and $n$ factors of $P$ is (the binomial coefficient ${m \choose n}$ picks a convenient normalization)
\begin{align}
\label{eq:DmnDef}
	D_{m,n} = {m+n \choose m}\, \int_{-1}^1 \frac{d \zeta }{2} \weyl{ X_\zeta^m P^n_\zeta }\,.
\end{align}
This can be lifted to a superfield $\SF{D}_{m,n}$. The ghost partners are then
\begin{align}
	[ \QSK , D_{m,n} ] &= m \,D_{m-1,n} \,\Psi + n \,D_{m,n-1}\, P_{\Psib} , \nonumber \\
	[ \QSKb , D_{m,n} ] &= m \,D_{m-1,n} \,\Psib + n\, D_{m,n-1}\, P_{\Psi} ,
\end{align}
while the top component is
\begin{align}
	\{ \QSKb , &[ \QSK, D_{m,n} ] \}  = {m+n \choose m}\, \left(  \weyl{X_\skR^m P_\skR^n }- \weyl{X_\skL^m P_\skL^n } \right)\nonumber \\
		&\qquad+ m\, (m-1)\, D_{m-2,n} \, \Psib \Psi + 
		mn \,D_{m-1,n-1}\, ( P_\Psi \Psi + \Psib P_{\Psib} + n\, (n-1) \,D_{m,n-2} \,P_\Psi P_{\Psib} )  .
\end{align}

\paragraph{3. The average multiplet:} 
This is the multiplet generated by the action of $\QSK, \QSKb$ on average operators. Continuing to focus on Weyl ordered operators, we construct the average super-operator\footnote{ For average operators it is easy enough to pass to the usual normal ordered basis, since the  super-operator $\frac{1}{2} \left(\SF{X}_{\zeta=1}^m \, \SF{P}_{\zeta=1}^m  + \SF{P}_{\zeta =-1}^n \, \SF{X}_{\zeta =-1}^m \right) $ gives the correct Schwinger-Keldysh average for the composite operator $\hat{X}^m \hat{P}^n$ in the single-copy theory, i.e., the symmetrization and order reversal can be carried out explicitly by hand.}
\begin{equation}
\begin{split}
\SF{\mathcal{A}}_k 
& =
	 \frac{1}{2} \left(\SF{X}_\zeta  + \alpha \, \SF{P}_\zeta \right)^k \bigg|_{\zeta =1} +  \frac{1}{2} \left(\SF{X}_\zeta  + \alpha \, \SF{P}_\zeta\right)^k \bigg|_{\zeta =-1}  \\
&=
	\mathcal{A}_k + \thb \, k\, \mathcal{A}_{k-1}\left(\Psi + \alpha\, P_{\Psib} \right)
	+ \theta\, k \, \mathcal{A}_{k-1}\left(\Psib + \alpha \, P_{\Psi} \right) + \thb\, \theta\, \widetilde{\mathcal{A}}_k \\ 
\mathcal{A}_k 	
&= 
	\frac{1}{2} \big[ \left(X_\skR + \alpha\, P_\skR\right)^k +  \left(X_\skL + \alpha\, P_\skL\right)^k \big] \\
\widetilde{\mathcal{A}}_k
&= 
	k \, \mathcal{A}_{k-1} ( \tilde X + \alpha \tilde P ) + k ( k-1) \mathcal{A}_{k-2} \left(\Psib \,\Psi + 
 	\alpha\,\left(P_{\Psi} \, \Psi + \Psib\, P_{\Psib} \right) + \alpha^2\, 	P_{\Psi}\, P_{\Psib} \right).
\end{split}
\label{eq:avmul}
\end{equation}

In components, the bottom component of an average multiplet with $m$ factors of $X$ and $n$ factors of $P$ is
\begin{align}
	A_{m,n} = \frac{1}{2}\,{m+n \choose m}\,  \left( \weyl{ X^m_R P^n_R } +  \weyl{ X^m_L P^n_L } \right). 
\label{eq:AmnDef}
\end{align}
The ghost partners are
\begin{align}
	[ \QSK , A_{m,n} ] &= m \,A_{m-1,n} \,\Psi + n \,A_{m,n-1} \,P_{\Psib} , \nonumber \\
	[ \QSKb , A_{m,n} ] &= m \,A_{m-1,n} \,\Psib + n \,A_{m,n-1}\, P_{\Psi} ,
\end{align}
while the top component is
\begin{align}\label{eq:averageTop}
	\{ \QSKb , &[ \QSK, A_{m,n} ] \}  = m\, A_{m-1,n} \,\tilde{X} + n \,A_{m,n-1} \,\tilde{P} \nonumber \\
		&\qquad+ m (m-1)\, A_{m-2,n}\, \Psib \Psi + mn\, A_{m-1,n-1} \,( P_\Psi \Psi + \Psib P_{\Psib} + n (n-1) A_{m,n-2} P_\Psi P_{\Psib} )  .
\end{align}

There are various other multiplets we could construct, but the last two will play a starring  role in our discussion. It is  instructive to note that not only are composite difference operators dressed with the ghost-bilinears as in 
\eqref{eq:Dtop}, but also that the average composite operator and the difference composite operator belong to different multiplets. So when we compute Schwinger-Keldysh Av-Dif correlators we should choose our super-operators accordingly.\footnote{ The discussion in Section 9 of \cite{Haehl:2016pec} ignores both these distinctions when deriving various constraints on super-operator correlators.  In our analysis below we will clarify some of the statements described therein. Structurally nothing really changes, but one has to account carefully for the above mentioned differences. }  

The ghost dressing comprises of three distinct combinations of bilinears
\begin{equation}
\begin{split}
\Psib\, \Psi \,, \qquad P_{\Psi} \, P_{\Psib} \,, \qquad P_\Psi \, \Psi + \Psib \, P_{\Psib} \,.
\end{split}
\label{eq:gdress}
\end{equation}
The  fourth ghost number zero combination $P_\Psi \, \Psi -\Psib \, P_{\Psib}$ does not enter into any ghost dressing. In itself this is an interesting statement, since a-priori it is not clear that the ghost dressings do not mess up the argument about difference operator correlators vanishing. The fact the one linear combination is unconstrained allows sufficient freedom to argue that the super-algebra structure described in \S\ref{sec:skreview} works as advertised.

\subsection{The super-embedding of states}
\label{sec:superem}

We have now assembled the machinery to implement the discussion of \S\ref{sec:require}. We would like to construct a super-embedding of an arbitrary initial state $\rhoi$ and find a final state satisfying the Conditions 1-3 outlined therein. We cannot get by with any super-embedding because as we have seen the operators (especially the difference operators) get dressed with ghost corrections. We have to ensure that these dressings do not spoil the physical requirement that we reproduce the Schwinger-Keldysh correlators, and work such that the ghosts decouple in the appropriate observables \eqref{eq:ghostdecoupling}. 

To achieve this we would like to set correlation functions involving ghost corrections to zero by suitably extending 
$ \kett{f_{SK} } $ and $\kett{\rho_{SK}}$ to states in the quadrupled Hilbert space $\kett{\SF{f}}$ and  $\kett{\SF{\rho}_{SK}}$. Passing into the creation/annihilation basis,  the ghost corrections are 
\begin{equation}
\begin{split}
2  \Psib\, \Psi(t) 
&=
	-b b^\dagger  + b c \,e^{2it}+ c^\dagger b^\dagger \,e^{-2it} - c^\dagger c ,  \\ 
2 P_\Psi P_{\Psib}(t) 
&=
	-b b^\dagger  - b  c \,e^{2it} - c^\dagger b^\dagger \,e^{-2it}-  c^\dagger c ,  \\ 
- i ( P_\Psi \Psi +  \Psib\, P_{\Psib} )(t) 
&= 
	c^\dagger b^\dagger   e^{2it} - b c \,e^{-2it} ,
\end{split}
\label{eq:gdressaad}
\end{equation}
where we are considering the Heisenberg operators in the extended Hilbert space 
$\mathcal{O}(t) = \mathscr{U}^\dagger(t) \, \mathcal{O}  \,\mathscr{U}(t)$. 

Firstly, the general implementation of \eqref{eq:fisuper} in the present context is given by (with a suitable normalization for the initial state)
\begin{align}
\kett{\SF{f}} &= \sum_{i,j,\alpha,\beta} f_{ij\alpha \beta} \, \kett{ij\alpha \beta},
&&\kett{\SF{\rho}} = \sum_{i,j,\alpha,\beta} \; \rho_{ij\alpha \beta} \, \kett{ij\alpha \beta}.
\label{eq:QHOfi}
\end{align}
We need to choose the coefficients $f_{mnij}$ and $\rho_{mnij}$ appropriately to ensure that we satisfy Conditions 1-3 and in the process ensure that \eqref{eq:gdressaad} are innocuous. 
The general solution to our requirements is\footnote{
One can check that the Conditions 2 for a final state $\kett{\SF{f}}$ of the form given in Eq.~\eqref{eq:QHOfi} gives 
\begin{equation*}
\begin{split}
f_{i,j,1,0} = \delta_{i,j} f_{i,1,0} \hspace{10mm}& f_{i,j,0,1} = \delta_{i,j} f_{i,0,1}  \\
f_{i,j,1,1} = \delta_{i+1,j} f_{i,1,1} \hspace{10mm}&  f_{i,j,0,0} = \delta_{i,j+1} f_{j,0,0}
\end{split} 
\end{equation*}
and the Condition 3 gives 
\begin{equation*}
\begin{split}
f_{i,0,1} = f_{i+1,0,1} \hspace{10mm} \sqrt{i+1} (f_{i,1,1} -f_{i,0,0}) = \sqrt i (f_{i-1,1,1} - f_{i-1,0,0}) \\
f_{i,1,0} = f_{i+1,1,0} \hspace{10mm} \sqrt{i+1} (f_{i,1,1} -f_{i,0,0}) = \sqrt i (f_{i-1,1,1} - f_{i-1,0,0})
\end{split}
\end{equation*} 
This is solved by $f_{i,0,1}=f_{0,1},f_{i,0,1}=f_{1,0}$ and $f_{i,1,1}=f_{i,0,0} +  \frac{1}{\sqrt{i+1}}$. }
{\small
\begin{equation}
\kett{\SF{f}}  =  \sum_{i}  \left( f_{1,0} \,  \kett{i i10} + f_{0,1}\,  \kett{i i 0 1}  + \left( f_{i,0,0}+ \frac{1}{\sqrt{i+1}} \right)  \kett{i\, (i+1)\, 1 1}  + f_{i,0,0} \, \kett{(i+1)\, i 0 0}  \right).
\end{equation}
}\normalsize

We however can get by without using all the freedom in the above solution. It suffices to simply pick a state that is 
the ground state in the ghost Hilbert space. The basic solution we will work with is simply
\begin{align}
\kett{\SF{f}} &= \sum_{i} \, \kett{ii10}\,,
&&\kett{\SF{\rho}} = \sum_{i,j} \; \rho_{ij} \, \kett{ij10}\,.
\label{eq:extensionChoice}
\end{align}
This extension has the following properties:
\begin{itemize}

\item The ghost corrections \eqref{eq:gdressaad} for one point functions vanish for all time: 
\begin{align}
\vev{ \Psib\, \Psi (t) } = \vev{ P_\Psi P_{\Psib} (t) } = \vev{ (P_\Psi \Psi +  \Psib\, P_{\Psib})(t)} = 0 \,.
\label{eq:ghostcorr1}
\end{align}
The correlation functions above are  the Schwinger-Keldysh observables, viz.,  $\vev{\mathcal{O}_1\, \cdots \mathcal{O}_n}
= \braa{\SF{f}} \mathcal{O}_1\, \cdots \mathcal{O}_n \kett{\SF{\rho}} =\braa{f_{_{SK}}} \mathcal{O}_1\, \cdots \mathcal{O}_n \kett{\rho_{_{SK}}}$ for purely bosonic operators. Eq.~\eqref{eq:ghostcorr1} ensures sure that 1-point functions of composite difference operators vanish, consistent with the Schwinger-Keldysh theory.

\item However, we can in fact  make a much more powerful statement: The ghost corrections annihilate the initial and final states!
\begin{align}
&\braa{\SF{f}}  \Psib\, \Psi = 0\,,
&& \Psib\, \Psi \, \kett{\SF{\rho}} = 0\,, \nonumber \\
&\braa{\SF{f}} P_\Psi P_{\Psib}= 0\,,
&&P_\Psi P_{\Psib} \, \kett{\SF{\rho}} = 0\,, \nonumber \\
&\braa{\SF{f}} ( P_\Psi \Psi +  \Psib\, P_{\Psib} ) = 0\,,
&&( P_\Psi \Psi +  \Psib\, P_{\Psib} ) \, \kett{\SF{\rho}} = 0\,.
\label{eq:gcorrect}
\end{align}
As a result we can always insert a ghost correction, making the difference operators constructed in 
\eqref{eq:difmul} $\QSK$ and $\QSKb$ exact, without altering correlators computed from the standard Schwinger-Keldysh construction. The only place where we have new behaviour is when we consider additional ghost operator insertions between the difference operator and both the initial and final states.

\item Importantly, correlation functions on the extended Hilbert space match those in the physical theory. For any string of operators picked from $\{\SKR{O}\,, \SKL{O}\}$ we have
\begin{equation}
\begin{split}
\vev{\{\SKR{O}, \SKL{O}\}} 
&= 
	\braa{f_{SK}} U_{SK}(T)  \,\{\SKR{O}, \SKL{O}\}\, U_{SK}(t) \kett{\rho_{SK}} \\
&= 
	\braa{ \SF{f}} \mathscr{U}(T) 
	\left( \{\SKR{O},\SKL{O}\} \otimes \Op{1}_{ghosts}\right)
	\mathscr{U}(t) \kett{\SF{\rho}} \,.
\end{split}
\end{equation}

\item While we have by construction ensured that the final state $\kett{\SF{f}}$ is $\QSK$ and $\QSKb$ closed, the same is not true about the initial state $\kett{\SF{\rho}}$. The background ghosts introduced in \cite{Haehl:2016pec} can be understood as the contributions obtained from $\QSK\kett{\SF{\rho}}$ and 
$\QSKb \kett{\SF{\rho}}$, respectively.
\end{itemize}

\subsection{Superspace correlators}
\label{sec:sliding}

We are now in a position to revisit the superspace constraints on correlation functions described in section 9 of \cite{Haehl:2016pec}. We remind the reader that their discussion assumed the existence of a suitable multiplet with bottom component being average and top component being a difference operator, which is only true at the level of fundamental operators. Furthermore, the underlying BRST supersymmetry relates correlators of objects within a single supermultiplet. These BRST Ward identities were easiest to derive in superspace, since there they follow from super-translation invariance. 

The super-correlators of interest are generic $n$-point functions of super-operators with suitable Schwinger-Keldysh ordering:
\begin{align}
\vev{\SF{{\cal T}}_{SK}\, \SF{\mathcal{O}}_1 \, \SF{\mathcal{O}}_2 \, \cdots \SF{\mathcal{O}}_n }
\equiv \vev{ \SF{{\cal T}}_{SK}\, \prod_{k=1}^n
\, \left( \mathcal{O}_k + \theta_k\, \mathcal{O}_{{\bar G},k} + \bar{\theta}_k \, \mathcal{O}_{G,k}+ \bar{\theta}_k\,\theta_k\, \widetilde{\mathcal{O}}_k \right)} \,.
\label{eq:GeneralSKScorr}
\end{align}
We have schematically indicated the structure of the super-operators in the definition above (more on this below).  Expanding the correlator in superspace  will lead to various terms involving the Grassmann coordinates $\bar{\theta}_i$ and $\theta_j$. Imposing super-translation invariance in these coordinates implies relations between these components of the correlator. A consistent set of solutions to such relations was found in \cite{Haehl:2016pec}, only upon inserting into the correlation function a background ghost operator,
\begin{equation}
\SF{\mathcal{O}}_0 = \mathbb{1} + \theta_0 \,  {\bar {\sf g}}_0  + \thetab_0 \,  {\sf g}_0 +  \thetab_0 \theta_0 \; {\sf d}_0 \,,
\label{}
\end{equation}
whose elements were interpreted as zero modes.
As presaged in \S\ref{sec:superem} this background ghost can be understood as arising from the super-embedding of the initial density matrix.  

 Non-vanishing correlators are those with vanishing ghost number, which provides a superselection rule. Furthermore, since the relations alluded to above relate terms with equal number of $\bar{\theta}_i\theta_j$ pairs, we break up the super-correlation function into {\it levels} based on the number of these pairs. Following \cite{Haehl:2016pec}, the set of  $n$-point correlation functions having $n_d$ pairs of $\bar{\theta}_i\theta_j$, is said to be at level $n_d$, and these are denoted as  
 ${}^n{\bf L}_{n_d}$. It was assumed that a correlator of type ${}^n{\bf L}_{n_d}$ contains at most $n_d$ difference fields. In the present context, the more precise statement is that it ${}^n{\bf L}_{n_d}$ contains correlators with at most $n_d$ top component fields (which need not be difference fields, depending on the supermultiplet used). We will continue to use this nomenclature to refer to correlation functions with at most $n_d$ difference operators (and any other operator as long as it is not $\QSK$ and $\QSKb$ exact).

In \cite{Haehl:2016pec} it was assumed that each super-operator has an average operator for the bottom component and the difference operator for its top component. We have seen that this is true only for the fundamental operators $\SF{X}$ and $\SF{P}$, but not for composite operators built out of these. In particular, we have to contend with two distinct multiplets $\SF{\mathcal{D}}_k$ and $\SF{\mathcal{A}}_k$ introduced in \eqref{eq:difmul} and \eqref{eq:avmul}, respectively, if we wish to talk about Schwinger-Keldysh average and difference operators. In addition, the top component of $\SF{\mathcal{D}}_k$ involves not just the difference operator of interest, but also its ghost dressing. 

To proceed let us record the schematic structure of our generic average and difference multiplets visually.
We have two different multiplets, and they are both involved when we derive the selection rules arising from super-translational invariance. To facilitate the discussion let us abstract the operators of interest as follows:
\begin{equation}
	\begin{tikzcd}
		&\avB \arrow{ld}{\QSK} \arrow{rd}[below]{\QSKb}    &   \\
		\ghA \arrow{rd}{\QSKb} & & \ghbA \arrow{ld}[above]{\!\!\!\!\!\!\!\!\!\!\!\!\!  - \QSK}\\
		&   \tilde{\avB} + \ggbA  &
	\end{tikzcd}
	~ \qquad \qquad 
	\begin{tikzcd}
		& \difB \arrow{ld}{\QSK} \arrow{rd}[below]{\QSKb}    &   \\
		\ghD \arrow{rd}{\QSKb} & & \ghbD \arrow{ld}[above]{\!\!\!\!\!\!\!\!\!\!\!\!\!  - \QSK}\\
		&   \difT + \ggbD  &
	\end{tikzcd} .
\label{eq:adgen}
\end{equation}
Each operator in these diagrams schematically stands for an infinite number of composite operators that may occur in the average and difference multiplets. To be explicit, we can take 
\begin{equation}
\avB \in \{ A_{m,n} \}\,, \qquad \difB \in \{ D_{m,n} \} 
\label{eq:}
\end{equation}	
as defined in \eqref{eq:AmnDef} and \eqref{eq:DmnDef}, respectively. The remaining fields in these diagrams are then as in \S\ref{sec:QHOcomp}.

We would like to show the following: {\it All correlation functions involving ghosts are determined in terms of standard Schwinger-Keldysh correlators of $\avB$ and $\difT$. Furthermore, these are consistent with the largest time equation, which says that difference operators cannot be future-most.}

Demonstrating this in full generality would require us to consider all possible multiplets (which may not be exhausted by the average and difference multiplets). We refrain from this task, and simply give an abstract argument that it holds true: we have in previous sections presented an explicit Hilbert space embedding. Any arbitrary correlation function can be computed using the rules of the previous sections, which manifestly implement the SK-BRST symmetries. Since super-translational invariance of super-correlators is equivalent to the action of the SK-BRST charges, the statement above must hold.

While this abstract argument is sufficient, it is more instructive to give some examples. To this end, we turn again to average and difference multiplets. Consider the super-correlation function 
\begin{align}
\vev{\SF{{\cal T}}_{SK}\, \SF{\mathcal{O}}_1 \, \SF{\mathcal{O}}_2 \, \cdots \SF{\mathcal{O}}_n \,\SF{\mathcal{O}}_0} 
\equiv \braa{\SF{f}} \SF{\mathcal{O}}_1 \, \SF{\mathcal{O}}_2 \, \cdots \SF{\mathcal{O}}_n  \kett{\SF{\rho}}\,.
\label{eq:ScorrTot}
\end{align}
The argument is structurally the same as the one given in \cite{Haehl:2016pec} with two main new ingredients: $(i)$ the background ghost operator $\SF{\mathcal{O}}_0$ is equivalent to a consistent superspace uplift of $\rhoi$, and $(ii)$ we will account for the structure of the average/difference composite operator multiplets in \eqref{eq:adgen}.

\paragraph{One-point functions:} The simplest analysis is for one-point functions (these were not considered in \cite{Haehl:2016pec}). One can either have an average or a difference multiplet and in either case the bottom component can have a non-vanishing expectation value depending on the initial state. The top components however would have to have vanishing expectation value; for the difference supermultiplet this embodies the largest time equation. Let us see how this works in turn, organizing the discussion by levels as described above.

\paragraph{$^1\mathbf L_0$:} The only correlators here are  $\vev{\avB}$ or $\vev{\difB}$ which are unconstrained since they do not contain any ghosts.
\paragraph{$^1\mathbf L_1$:} These are correlators containing one difference operator. The largest time equation demands that 
$\vev{\difT} =0$. However, we find that the top-component of the difference multiplet for composite operators is not simply $\difT$ but rather it gets dressed with ghost bilinears. Superspace Ward identities only can demand that $\QSK$ and $\QSKb$ exact operators have vanishing correlators, so we are only free to conclude that $\vev{\difT + \ggbD} =0$. We can draw two conclusions from this. First, using the boundary conditions \eqref{eq:gcorrect}, we infer the largest time equation:
{\small
\begin{equation}
\begin{split}
0 &= \vev{\difT + \ggbD} \\ 
 &= \vev{\difT} + \braa{ \SF{f}} m\, (m-1)\, D_{m-2,n} \, \Psib \Psi + 
		mn \,D_{m-1,n-1}\, ( P_\Psi \Psi + \Psib P_{\Psib} + n\, (n-1) \,D_{m,n-2} \,P_\Psi P_{\Psib} )\kett{\SF{\rho}} \\
&\!\!\! \stackrel{\text{\tiny \eqref{eq:gcorrect}}}{=} \;\;\vev{\difT}  \,.
\label{eq:L11lt}
\end{split}
\end{equation}	
}\normalsize
for any value of $k$.

Second, we can extract the complete set of constraints on the ghost correlators at this level. To this end, we start with 
\begin{align}
& 0 = \vev{ \difT +  \ggbD} 
	= \braa{\SF{f}} \{ \QSKb , \ghD \} \kett{\SF{\rho}} 
	= - \braa{\SF{f}} \{ \QSK , \ghbD \} \kett{\SF{\rho}} \,.
\end{align}
We can then use fact that the final state is annihilated by $\QSK$ and $\QSKb$ to infer that 
\begin{equation}
\begin{split}
& 0 =  \braa{\SF{f}} \ghD\, \QSKb \kett{\SF{\rho}} = \braa{\SF{f}} \, \ghbD\, \QSK \kett{\SF{\rho}}  \\
&
\;\; \Longrightarrow \;\;
0 = \vev{\ghD\, \bar{\sf g}_0 } = \vev{\ghbD\, {\sf g}_0} 
\end{split}
\label{eq:g0corr1}
\end{equation}
where in the last step we identify 
\begin{equation}
\QSKb \kett{\SF{\rho}} = \bar{\sf g}_0 \kett{\SF{\rho}}  \,, \qquad \QSK \kett{\SF{\rho}} = {\sf g}_0 \kett{\SF{\rho}}  \,.
\label{eq:g0modes}
\end{equation}	
We now see explicitly that the ghost zero modes of \cite{Haehl:2016pec} are simply insertions of the Schwinger-Keldysh supercharges themselves.\footnote{  In \eqref{eq:g0modes} the ghost zero modes are indicated as operators that act on the extended Hilbert space. Equivalently, one can view them as the Grassmann-odd ghost partners of the density matrix as can be seen from the first line of \eqref{eq:g0corr1}. }

\paragraph{Two-point functions:} Let us look at two point functions where the operators are inserted at times $t$ and $t'$ respectively, and we suppose for definiteness that  $t'>t$. 

\paragraph{$^2\mathbf L_0$:} This set contains correlators of the form $\vev{ \avB(t') \,\avB(t) }$ or other bottom components. These correlation functions do neither involve any difference operators nor ghosts and consequently there is nothing to be constrained. 

\paragraph{$^2\mathbf L_1$:} At level one we have the insertion of a single difference operator. This can either be at $t'$ or at $t$, and we treat these in turn (thus being explicit about the Schwinger-Keldysh time-ordering).

Let us  first examine correlators with the difference operator at the largest time $t'$. By commuting the supercharges across the operator insertions (which is equivalent to imposing super-translational invariance) we find:
\begin{equation}
\begin{split}
	\vev{ (\difT' + \ggbD') \, \avB } &= \vev{ \{ \QSKb , \ghD' \}  \avB } = \vev{ \ghD' \, \QSKb \avB } 
		= \vev{ \ghD' \left( [\QSKb , \avB ] + \avB \QSKb \right)}  \\
& = 
	\vev{ \ghD' \ghbA } + \vev{ \ghD'  \,\avB \,\bar{\sf g}_0 } 
	= \braa{\SF{f}} \ghD' \,\ghbA + \ghD'  \,\avB \,\QSKb \kett{\SF{\rho}} \,.
\label{eq:slidingRule1}
\end{split}
\end{equation}
Similarly, we derive a second identity using the fact that the top component of the difference multiplet is $\QSK$-exact, viz., 
\begin{align}
 \vev{ \ghbD' \ghA } + \vev{ \ghbD'  \,\avB \,{\sf g}_0 } 
	= \braa{\SF{f}} \ghbD' \, \ghA + \ghbD'  \,\avB \,\QSK \kett{\SF{\rho}} \,.
\end{align}

If we  consider the average operator to be at the largest time, we derive instead: 
\begin{equation}
\begin{split}
\vev{\avB' \, \difT}  + \vev{\ghbA'\, \ghD} 
& = 
	\braa{\SF{f}}\avB' \, \ghD\, \QSKb\kett{\SF{\rho}} \ \\
\vev{\avB' \, \difT}  + \vev{\ghA'\, \ghbD}
& = 
	 + \braa{\SF{f}}\avB' \, \ghbD\, \QSK\kett{\SF{\rho}} \ \\
\end{split}
\label{eq:}
\end{equation}
These are the equations obtained from super-translational invariance in 
\cite{Haehl:2016pec}, cf., the first two equations of Eq.~$(9.8)_{_\text{\cite{Haehl:2016pec}}}$. The other two equations can be similarly derived and one finds similar expressions with ${\sf d}_0 \kett{\SF{\rho}} \equiv \QSKb \QSK \kett{\SF{\rho}}$: 
\begin{equation}
\begin{split}
\vev{ \avB'\, \avB {\sf d}_0 } 
&= 
	- \vev{ \avB' \, \ghbA {\sf g}_0} - \vev{\ghbA' \, \avB \, {\sf g}_0}  \\
\vev{ \avB'\, \avB {\sf d}_0 } 
&= 
	- \vev{ \avB' \, \ghA\, \bar{\sf g}_0} - \vev{ \ghA' \, \avB\, \bar{\sf g}_0 } .
\end{split}
\label{eq:}
\end{equation}
This determines all average-difference-ghost correlators at this level in terms of standard Schwinger-Keldysh correlators.

One can go further by using the boundary conditions \eqref{eq:gcorrect}. For instance, these imply that $\vev{ (\difT' + \ggbD') \, \avB }=\vev{ \difT'  \, \avB }$, which vanishes if $t'>t$ due to the largest time equation.

\paragraph{$^2\mathbf L_2$:} Finally, consider two-point functions with two difference operators and suitable ghost dressing. We now have to modify the statements in \cite{Haehl:2016pec}. For instance,  Eq.~$(9.10)_{_\text{\cite{Haehl:2016pec}}}$ of that paper is modified by ghost corrections $\ggbD$. This happens because  the ghost corrections appear sandwiched in between two ghost operators and so cannot annihilate the initial or final state. 

To be self-contained, let us start with  $\vev{ \difT' \difT }= 0$ and using the superalgebra one furthermore finds
\begin{equation}
\begin{split}
0 &= 
	\vev{ \ghD'\,\left(\difT + \ggbD\right)\,  \bar{\sf g}_0}  \,, \\
0 &= 
	\vev{ \ghbD\,\left( \difT  + \ggbD\right)\, {\sf g}_0}  \, .
\end{split}
\end{equation}

It is easy to continue with this analysis for higher-point functions. The ingredients are always: largest-time equation, boundary conditions, and super-translational invariance. Up to some small adjustments to take care of the ghost dressing the basic story outlined in \cite{Haehl:2016pec} carries through. One finds precisely the same constraints as in \cite{Haehl:2016pec}, the only modification being that average, difference and ghost operators get replaced by the respective components of either average or difference multiplets. In fact, one obtains two sets of relations, each isomorphic to those in \cite{Haehl:2016pec}: those for the average, and those for the difference multiplet. 

It is instructive that we have now given a physical picture for the background  ghost insertion in the correlation function \eqref{eq:gcorrect}. The reader can convince themselves that the total number of relations obtained from super-translational invariance is identical to that described in \cite{Haehl:2016pec}, though now the relation does not quite set certain correlators to zero, but rather fixes them in terms of some other correlation function. Furthermore, the ghost bilinear  $P_\Psi \, \Psi -\Psib \, P_{\Psib}$ which did not appear in any of the dressings discussed here, but it has its correlators determined by the explicit super-embedding.

\paragraph{Example:} For illustration, let us check (\ref{eq:slidingRule1}) in a particular case. When the difference operator is $X_\skR - X_\skL$ and the average operator is $\frac 1 2 ( P_\skR + P_\skL )$, we have $\ghD' = \Psi$, $\ghbA = P_\Psi$. Equation (\ref{eq:slidingRule1}) then reads
\begin{equation} \label{eq:slidingRuleEx}
 0 = \braa{\SF{f}} \Psi P_\Psi + \frac 1 2  (P_\skR + P_\skL) \Psi \QSKb \kett{\SF{\rho}} \,.
\end{equation}
For simplicity we have taken both operator insertions to be at $t=0$ but with the difference operator placed to the left of the average operator.

If the system is in the ground state $\left |0\right \rangle$ of the quantum harmonic oscillator, the extension (\ref{eq:extensionChoice}) gives
\begin{align}
	\kett{\SF{\rho}} = \kett{ 0 0 1 0 },
	&&\kett{ \SF f } = \sum_i \kett{ ii 1 0 }.
\end{align}
It's then a simple matter to explicitly compute the expectation values using (\ref{eq:raisingLoweringDef}) and (\ref{eq:Qaad})
\begin{align}
	\braa{\SF{f}}\Psi P_\Psi\kett{\SF{\rho}} = i,
	&&\frac 1 2 \braa{\SF{f}} (P_\skR + P_\skL) \Psi \QSKb \kett{\SF{\rho}} = - i .
\end{align}
We see that the relation (\ref{eq:slidingRuleEx}) is satisfied.

\section{Generalizations \& open questions}
\label{sec:discuss}

In this work, we have addressed certain subtleties in the doubling of Hilbert space and the operator structure of ghosts associated with the Schwinger-Keldysh construction, which were not evident in the abstract analysis of \cite{Haehl:2016pec}. Our discussion was facilitated by moving from a general path integral description to a Hilbert space based construction in the simplest of quantum models: the quantum harmonic oscillator. This also allowed us to demonstrate in this elementary setup various general claims relating to the structure of Schwinger-Keldysh theories.

The main features missing from the earlier discussions are as follows:
\begin{enumerate}
\item Ghost dressing of composite difference operators by ghost bilinears. Here, using an explicit construction,
we are able to describe how this dressing works for an arbitrary composite operator in the quantum oscillator.
\item Ghost partners of the density matrix (also referred to as ghost zero modes) playing a crucial role. This was
assumed without derivation in the aforementioned previous work. Here, we can explicitly construct and confirm the picture
posited before.
\item Final state boundary condition. We have shown that there exists a final state which is annihilated by the BRST charges which provides an appropriate future boundary condition for the ghost fields. 
\end{enumerate}
With these two ideas taken into account, we have demonstrated that the full operator structure of the quantum oscillator can be embedded within the superspace formalism. Given that perturbative QFTs can be recast into deformations of a theory of infinitely many quantum oscillators, we expect our discussion to carry through to them in a straightforward way, as we now argue, before turning to some general lessons.

\subsection{Interacting theories}
\label{sec:interact}

Let us first see that we can straightforwardly add interactions to our quantum mechanics model.
For concreteness, we can add a quartic interaction $H_{\text{int}} = \frac{\lambda}{4!} \OpH{X}^4$, though all of the discussion applies equally well to any more general interaction. 

Fortunately, the formalism developed above is readily adapted. The uplift to the quadrupled Hilbert space works in exactly the same way, with the same supercharges given in (\ref{eq:QactXP}) and the same $\kett{\SF{f}}$. As before, we demand that the super-extension satisfies three requirements. Since we have not modified the supercharges, $\kett{\SF{f}}$ remains in the kernel of the $\QSK$ and $\QSKb$. We also require that $\kett{\SF{f}}$ is a zero energy eigenstate of the extended Hamiltonian $\mathscr{H}$, but to do so we must first determine $\mathscr{H}$ for the interacting system. As discussed extensively above, the naive difference operators for composite operators, such as our interaction term $\OpH{X}^4$, must be dressed with ghost corrections in order to be $\QSK$ and $\QSKb$ exact. Therefore, if the (extended) Hamiltonian is to be $\QSK$ and $\QSKb$ exact, then we must include these ghost corrections in $\mathscr{H}$: 
\begin{equation}   \label{eq:H-interacting}   \mathscr{H} = \mathscr{H}_0 +
\frac{\lambda}{4!} (X_\skR^4 - X_\skL^4) + \frac{\lambda}{2} D_{2,0} \, \Psib
\Psi    = \mathscr{H}_0 + \frac{\lambda}{4!} (X_\skR^4 - X_\skL^4) +
\frac{\lambda}{2} \left(X^2 + \frac{1}{12} \tilde{X}^2 \right) \Psib \Psi\, ,
\end{equation}
using the notation of \eqref{eq:DmnDef}.
With this extended Hamiltonian and our choice of $\kett{\SF{f}}$ in (\ref{eq:extensionChoice}), it is straightforward to check that $\mathscr{H} \kett{\SF{f}} = 0$. Note that superspace is designed to compactify notation: in the above example, we can simply write $\mathscr{H} = \mathscr{H}_0 + \int d\theta d\thb \, \frac{\lambda}{4!}\,\SF{D}_{4,0}$.

The argument readily extends to general interactions; we just note that with our choice of final state \eqref{eq:extensionChoice}, $\Psi \kett{\SF{f}} = P_{\Psib}  \kett{\SF{f}} = 0$. Therefore taking  $\mathscr{H}_{int} = \{\QSKb, [\QSK, D_{m,n}]\} $, we have
  \begin{equation*}
    \mathscr{H}_{int} \kett{\SF{f}} = \QSKb [\QSK, D_{m,n}] \kett{\SF{f}} = \QSKb \left( m D_{m-1,n} \Psi + n D_{m,n-1} P_{\Psib} \right) \kett{\SF{f}} = 0\, .
  \end{equation*}
Superspace again allows for compact notation: $\mathscr{H}_{int} = \int d\theta d\thb \, \SF{D}_{m,n}$.

Finally, we need to check that the extension doesn't modify any of the purely bosonic correlators. The only place the interactions could modify this condition is via the ghost terms in the Hamiltonian that will arise from unitaries implementing time evolution. However, since these ghost corrections annihilate our final state, cf.,  Eq.~\eqref{eq:gcorrect}, the same argument as in the non-interacting case applies. Thus, the addition of interactions doesn't pose any obstacles to lifting the theory to the extended Hilbert space and the formalism applies for generic Hamiltonians.

\subsection{Lessons for QFTs}
\label{sec:lessons}

The main rationale behind undertaking the exercise of analyzing the Hilbert space picture of Schwinger-Keldysh formalism and the postulated BRST symmetries inherent therein is to get a clearer picture in favour of the abstract arguments described in 
\cite{Haehl:2016pec}. We now turn to ask: how far can we take the lessons from the quantum mechanics picture?

We will now attempt an  argument directly at the level of the operator algebra, notwithstanding the fact that beyond quantum mechanics we have to confront an abstract structure which may not necessarily be generated in terms of simple fundamental building blocks (perturbative field theories can be dealt with as in \S\ref{sec:QHO} and \S\ref{sec:interact}). 
The basic hypotheses underlying our construction are: (a) the existence of a pair of BRST symmetries $\{\QSK,\QSKb\}$ whose action on the doubled Schwinger-Keldysh operator algebra is well defined, and (b) this action can be uplifted into a 
graded Hilbert space with suitable ghosts appended. 

With these assumptions, we are left with testing the Ward identities arising from the BRST symmetries. For one, we want to ensure that the Schwinger-Keldysh difference operators can be suitably dressed with ghost bilinears to lie in the BRST cohomology. It is clear however that the BRST cohomology will contain other elements as well; for instance the average super-multiplet's top component has a bosonic part which is not necessarily a Schwinger-Keldysh difference operator. Determining the full set of such operators amounts to solving for the BRST cohomology, a well defined problem, given the operator algebra and the action of 
$\{\QSK,\QSKb\}$. 
 
There are two issues about this structure that are worth elaborating upon:
\begin{enumerate}
\item The BRST Ward identities arising from correlation functions involving top components of super-multiplets must be self-consistent. They should not constrain the structure of the Schwinger-Keldysh theory beyond correlators involving ghosts. This can of course be checked to hold given an explicit construction (as carried about above), but it seems implausible that it can fail once it is shown that the BRST cohomology is well-defined.
\item Additionally, we can ask if all the correlation functions of the extended theory are determined in terms of standard Schwinger-Keldysh correlators. We may not expect this to be unambiguous, since there is some freedom in the choice of the embedding of the initial state (and the final state, though that we expect leads to no additional freedom in observables). We believe it should be possible to fix all the correlation functions for 
the ghost sector in terms of usual Schwinger-Keldysh correlators, i.e., we anticipate that the discussion of \cite{Haehl:2016pec} will carry through with the new elements of ghost dressing discussed in \S\ref{sec:sliding}. The only part for which we have not yet given a clean argument are the ghost bilinears that do not appear as the dressing of any double copy operator. For example, in the harmonic oscillator we did not encounter $P_\Psi \, \Psi -\Psib \, P_{\Psib}$ as the ghost dressing for any operator. In this case we can however appeal to our explicit Hilbert space embedding to determine its correlations. How this structure extends more generally is a question that is worth investigating further. 
\end{enumerate}

As mentioned earlier, the aforementioned issues should not be a problem for perturbative QFTs. Here we may readily employ the same strategy as in our quantum mechanics example. Moreover, discussions of operator ordering etc., will be moot if we only consider, as is often the case, interaction terms which are built out of fields alone (and not their conjugate momenta). 

An interesting example for future analysis would be to understand non-perturbative interacting theories (say 2d minimal model CFTs) in this framework. The challenge here is to embed the standard OPE structure by adding ghost operators, ghost bilinear operators, and appropriate difference operator dressing by ghost bilinears. One would like to check whether ghost bilinear sectors and dressing can be made consistent with the closure of OPE. This appears to be a concrete setting where we can hope to get a handle on the questions raised above. 

Additional motivation for such an analysis comes from the idea of seeing whether the Schwinger-Keldysh superspace is a useful way of thinking about unitarity in these theories. Progress in this direction could potentially  also prove useful in giving a simple proof of the unitarity in superstring field theory, cf.,  \cite{Pius:2016jsl} for the existing proof. Another direction which goes to the heart of the reason for the introduction of these BRST symmetries and ghosts is in applications to open quantum systems and effective non-unitary field theories arising out of coarse-graining. It would especially be interesting to address the implications of our analysis for open $\phi^4$ theory (see the discussion in \cite{Avinash:2017asn}) and how in general open quantum systems are to be embedded within superspace. 

\acknowledgments

It is a great pleasure to thank Simon Caron-Huot, Veronika Hubeny, Chandan Jana, and Arnab Rudra for useful discussions. 
MC, DR, and MR are supported in part by U.S.\ Department of Energy grant DE-SC0009999 and by the 
University of California.  FH gratefully acknowledges support through a fellowship by the Simons Collaboration `It from Qubit'. RL gratefully acknowledges support from International Centre for Theoretical Sciences (ICTS), Tata Institute of Fundamental Research, Bengaluru.
FH would like to thank UC Santa Barbara, UC London, and McGill University for hospitality during various stages of this project.
MR would like to thank Tata Institute of Fundamental Research, Mumbai and International Centre for Theoretical Sciences, Bengaluru for hospitality during the concluding stages of this project.
PN and RL would  also like to acknowledge their debt to the people of India for their steady and generous support to research in the basic sciences. 

\appendix
\section{The ghost Hilbert space}
\label{app:InnerProduct}

In this appendix, we construct the inner product on the ghost Hilbert space. In the main text, we demand the Hermiticity condition
\begin{align}
	  \Psi^\dagger =   - { \bar \Psi} .
\end{align}
While other choices are possible, this is convenient as it is equivalent to
\begin{align}
	Q^\dagger = \bar Q .
\end{align}
Consistency with the Schrodinger equation for the Heisenberg picture ghost momenta
\begin{align}
	  P_\Psi = \partial_t   { \bar \Psi } = - i [  {\mathcal H} ,   { \bar \Psi} ],
	&&  P_{\bar \Psi} =  \partial_t   \Psi =- i [   {\mathcal H} ,   \Psi ]
\end{align}
then implies that
\begin{align}
	  P_{\Psi}^\dagger = i [   {\mathcal H} ,   \Psi ] =  - P_{\bar \Psi} 
\end{align}
as well.
With this choice, and using the definitions of the $b$, $c$ ghosts in (\ref{eq:raisingLoweringDef}) we then find that $b^\dagger$ and $c^\dagger$ truly are the hermitian conjugates of $b$ and $c$ respectively, so there is no inconsistency of notation.

We can now construct the inner product that respects these Hermiticity conditions. In the $b$ sector
\begin{align}
	&\langle 1 | 1 \rangle = \langle 0 | b b^\dagger | 0 \rangle = \langle 0 | \{ b , b^\dagger \} | 0 \rangle = \langle 0 | 0 \rangle ,
	\nonumber \\
	&\langle 1 | 0 \rangle = \langle 0 | b | 0 \rangle = 0 , \nonumber \\
	& \langle 0 | 1 \rangle = \langle 0 | b^\dagger | 0 \rangle = 0 ,
\end{align}
fixing the inner product up to a constant. We choose $\langle \alpha | \beta \rangle = \delta_{ij}$.
In the $c$ sector the same manipulations give
\begin{align}
	&\langle 1 | 1 \rangle = \langle 0 | c c^\dagger | 0 \rangle = \langle 0 | \{ c , c^\dagger \} | 0 \rangle = - \langle 0 | 0 \rangle ,
	\nonumber \\
	&\langle 1 | 0 \rangle = \langle 0 | c | 0 \rangle = 0 , \nonumber \\
	& \langle 0 | 1 \rangle = \langle 0 | c^\dagger | 0 \rangle = 0 ,
\end{align}
determining (up to a constant), that $\langle \alpha | \beta \rangle = \eta_{\alpha \beta} = \text{diag} (1,-1)$.

For the states $| \alpha \beta \rangle$, we have
\begin{align}\label{innerProduct}
	\langle \alpha \beta | \gamma \delta \rangle = \delta_{\alpha \gamma} \eta_{\beta \delta} =
	\begin{pmatrix}
		1 & 0 & 0 & 0 \\
		0 & -1 & 0 & 0 \\
		0 & 0 & 1 & 0 \\
		0 & 0 & 0 & -1 
	\end{pmatrix},
\end{align}
states being presented in the matrix from left to right and top to bottom as $| 00 \rangle, | 0 1 \rangle, | 1 0 \rangle , | 11 \rangle$.



\providecommand{\href}[2]{#2}\begingroup\raggedright\endgroup

\end{document}